\documentclass[11pt]{article}
\usepackage[bookmarks=false,colorlinks=true,linkcolor={blue}]{hyperref} 
\usepackage{epsfig}
\usepackage{amsmath}    
\usepackage{amssymb}
\usepackage{graphicx}
\usepackage{amsfonts}   
\usepackage{latexsym}   
\usepackage{slashed}
\usepackage{yfonts} 
\usepackage{verbatim}
\usepackage{wrapfig}
\usepackage{mathrsfs}
\usepackage{cancel}

\global\newcount\itemno \global\itemno=0

\def\itemaut#1{\global\advance\itemno by1\noindent\item{\the\itemno.}#1}

\newif{\ifeq}		
\eqfalse		
\newcommand{\be}{\begin{equation}}
\newcommand{\ee}{\end{equation}}
\newcommand{\bes}{\begin{equation*}}
\newcommand{\ees}{\end{equation*}}
\newcommand{\bea}{\begin{eqnarray}}
\newcommand{\eea}{\end{eqnarray}}
\newcommand{\beas}{\begin{eqnarray*}}
\newcommand{\eeas}{\end{eqnarray*}}

\def\({\left(}
\def\){\right)}
\def\[{\left[}
\def\]{\right]}
\def\<{\langle}
\def\>{\rangle}

\def\frac#1#2{{#1 \over #2}}

\def\vev#1{\langle{#1}\rangle}

\renewcommand{\a}{\alpha}

\newcommand{\f}{\psi}

\newcommand{\w}{\omega}
\newcommand{\m}{\mu}

\newcommand{\G}{\Gamma}

\newcommand{\U}{\Upsilon}

\newcommand{\GB}{\overline{\Gamma}}

\newcommand{\CA}{{\cal A}}

\newcommand{\CC}{{\cal C}}
\newcommand{\CD}{{\cal D}}

\newcommand{\CJ}{{\cal J}}

\newcommand{\CL}{{\cal L}}

\newcommand{\CN}{{\cal N}}
\newcommand{\CO}{{\cal O}}

\newcommand{\CQ}{{\cal Q}}

\newcommand{\CT}{{\cal T}}
\newcommand{\CU}{{\cal U}}
\newcommand{\CV}{{\cal V}}

\def\eg{{\it e.g.}}

\newcommand{\lsim}{\,\raise.3ex\hbox{$<$\kern-.75em\lower1ex\hbox{$\sim$}}\,}
\newcommand{\gsim}{\,\raise.3ex\hbox{$>$\kern-.75em\lower1ex\hbox{$\sim$}}\,}

\def\p{\partial}

\newcommand{\tp}{{\theta^+}}
\newcommand{\tbp}{{\bar{\theta}^+}}

\newcommand{\DB}{\overline{\rm D}}

\newcommand{\Dp}{{\rm D}_{+}}
\newcommand{\DBp}{\overline{\rm D}_{+}}

\newcommand{\QPB}{\overline{Q}_+}

\def\Ka{K\"{a}hler}

\def\nK{non-K\"{a}hler}
\def\TT{$(2,2)$}
\def\ZT{$(0,2)$}

\def\II{\relax{I\kern-.10em I}}

\usepackage{amssymb}

\numberwithin{equation}{section}
\usepackage{allanamscd}

\def\L{\Lambda}

\newcommand{\Dm}{{\mathcal D}_{-}}
\renewcommand{\f}{\frac}

\begin{document}

\begin{titlepage}
\begin{flushright}MIT-CTP-4377 \end{flushright}
\vskip 1in

\begin{center}
{\Large{GLSMs for \nK\ Geometries}}
\vskip 0.5in Allan Adams, Ethan Dyer, and Jaehoon Lee
\vskip 0.2in {\it Center for Theoretical Physics\\  Massachusetts Institute of Technology\\ Cambridge, MA  02139, USA}
\end{center}
\vskip 0.5in

\begin{abstract}\noindent
We identify a simple mechanism by which $H$-flux satisfying the modified Bianchi identity arises in garden-variety \ZT\ gauged linear sigma models.  
Taking suitable limits leads to effective gauged linear sigma models with Green-Schwarz anomaly cancellation.
We test the quantum-consistency of a class of such effective theories by 
constructing an off-shell superconformal algebra, identifying unexpected topological constraints on the existence of this algebra and providing evidence that these models run to good CFTs in the deep IR when all of the constraints are satisfied.  

\end{abstract}

\end{titlepage}

%
%
%
\section{Introduction}\label{Sec:Intro}

String theory has taught us a great deal about the quantum geometry of Calabi-Yau (CY) manifolds.
Central to this progress is the gauged linear sigma model (GLSM), a formalism which translates quantum computations in \Ka\ geometries into free-field-theory calculations in an auxiliary gauge theory \cite{Witten:1993}.  
Of course,  CYs form a set of measure zero in the full space of string compactifications, so it is natural to wonder what we can say about the quantum geometry of more general non-CY manifolds.  
This question is particularly natural in the heterotic string, where a worldsheet analysis should suffice.
This motivates us to search for GLSMs for non-K\"ahler\footnote{The term ``balanced'' is probably more appropriate \cite{Michelsohn:1982}, since all such 4d $\CN=1$ compactifications come from balanced manifolds which may or may not be \Ka.  The term ``non-K\"ahler'' has become standard, however, emphasizing that these manifolds need not be K\"ahler.} manifolds, with the goal of using them to study the quantum geometry of more general geometries.

At first glance, this seems quite challenging.  Mathematically, the basic structure of a GLSM is a \Ka\ quotient of flat space, which naively should not be much help in getting a \nK\ manifold.   Meanwhile, if the geometry is not \Ka, target space SUSY requires non-trivial 
3-form torsion, $H$, which must satisfy the Green-Schwarz (GS) Bianchi identity,
\be\label{EQ:BI}
dH = \a'\( {\rm tr} R\wedge R -  {\rm Tr} F\wedge F \)\,.
\ee
Correspondingly, the $B$-field itself must transform non-trivially under the full set of spacetime gauge transformations.   
Furthermore, $dH\neq0$  implies a non-trivial dilaton profile \cite{Strominger:1986}, so that the worldsheet conformal symmetry is only realized non linearly.  
Finally, the fact that tree- and one-loop effects compete in (\ref{EQ:BI}) means that some cycles may be frozen near string scale, making a large-radius limit problematic.  It is difficult to see how all these effects could be incorporated into a GLSM.  Indeed, considerable effort has been devoted to adding bells and whistles to the GLSM to mock these effects up \cite{Adams:2006, AdamsLapan:2009, AdamsGuarrera:2009, Adams:2009, Blaszczyk:2011, Quigley:2011}.

In this paper, we demonstrate a general mechanism for generating the modified Bianchi identity and related quantum effects in a garden-variety \ZT\ GLSM.\footnote{For useful introductions and reviews of (0,2) sigma models see \cite{Distler:1995,McOrist:2010}.}
Surprisingly, this mechanism does not require any new ingredients.  Rather, simple quantum effects in every \ZT\ GLSM generate all the necessary features dynamically.  
The basic mechanism first appeared in the study of 4d chiral gauge theories \cite{D'Hoker:1984,D'HokerFarhi:1984}.  In a chiral gauge theory, integrating out heavy fermions in chiral representations of the gauge group will generically generate anomalous Green-Schwarz terms in the action.  These terms are essential for canceling the anomaly of the surviving light fermions.  In our \ZT\ GLSMs, such GS terms precisely generate the corrections to the $B$-field transformation law which are required to satisfy the Bianchi identity, and which lead to non-trivial $H$-flux and \nK\ hermitian metric.

The central claim of this paper is that all of the features of a generic \nK\ model with $H$-flux can in fact be found within a standard \ZT\ GLSM.  As we shall see, non-trivial $H$-flux and a modified transformation law for the $B$-field are automatically generated as needed by the mechanism sketched above.  Indeed, we will find that all of the various previously-preposed quasi-linear mechanisms for generating $H$-flux in special cases emerge naturally when studying the low energy effective physics of standard \ZT\ GLSMs.

The rest of this paper is organized as follows.  In Section 2, we demonstrate how the GS mechanism in a GLSM generates precisely the anomalous transformation law for the $B$-field needed to ensure cancellation of the sigma model anomalies of the IR NLSM.  Along the way, we demonstrate that the GS models previously constructed can thus be UV completed into completely pedestrian \ZT\ GLSMs; conversely, at low energies and at generic points in the \ZT\ moduli space, a general \ZT\ GLSM reduces to such an anomalous GS effective GLSM.   In Section 3, we will study a particularly simple class of such effective GLSMs in which the axial couplings are entirely linear, and use them to explore the quantum consistency and semi-classical geometry of such GLSMs. Furthermore, we find unanticipated topological constraints on the existence of such models. 
The central ingredient in these models is a set of Green-Schwarz axions, $Y$,  playing the role of St\"uckelberg fields for the anomalous gauge symmetries.  
This allows us to avoid the subtleties associated with logarithms and address both classical and quantum properties of the models.  As we shall see, these theories show every sign of running to good IR CFTs -- more precisely, these theories enjoy a \ZT\  superconformal algebra which closes in $\QPB$-cohomology, as was previously shown for the $T^{2}$ models in \cite{AdamsLapan:2009}. 
While the models on which we focus are not generic, the lessons we learn can be readily applied to more general GLSMs with non-trivial $H$-flux. We close in Section 4 with a summary and list of future directions.

%
%
%

\section{Generating $dH$ in a \ZT\ GLSM}\label{Sec:Models}

The goal of the present section is to demonstrate that all the ingredients necessary to study models with non-trivial $dH$ are already present in a garden-variety \ZT\ GLSM.  We begin by recalling how $dH\neq0$ and the modified Bianchi identity arise in a general \ZT\ NLSM.  We then review how these effect can be incorporated into quasi-linear \ZT\ GLSMs by hand via Green-Schwarz anomaly cancellation, as first demonstrated in \cite{Adams:2006,Adams:2009}.  We then argue that such quasi-linear models arise as effective descriptions of totally standard (and non-anomalous) \ZT\ GLSMs at generic points in their moduli space.  Concretely, moving along the (0,2) moduli space modifies masses for chirally-gauged fermions in the GLSM; integrating out the heaviest fermions then generates non-linear Green-Schwarz terms in the action which realize the non-trivial $dH$.   This explains for example how $dH\neq0$ is generated in \ZT\ deformations of \TT\ GLSMs, such as deformations of the tangent bundle.  We also comment briefly on the effective geometry of such models. 


\subsection{Torsion in \ZT\ NLSMs}

Before we address the GLSM, let's recall how the modified Bianchi identity and the $c_{2}$ constraints arise in a non-linear sigma model (NLSM) with \ZT\ supersymmetry.  The action for a \ZT\ NLSM is given by \cite{Nibbelink:2012}:
\begin{eqnarray}
\mathcal{L}&=&\frac{1}{2}\left(G_{I\bar{J}}+B_{I\bar{J}}\right)\partial_{+}\phi^{I}\partial_{-}\bar{\phi}^{\bar{J}}+\frac{1}{2}\left(G_{I\bar{J}}-B_{I\bar{J}}\right)\partial_{-}\phi^{I}\partial_{+}\bar{\phi}^{\bar{J}}\nonumber\\
&&+\frac{i}{2}\left(\bar{\psi}_{+}^{\bar{J}}G_{I\bar{J}}(\partial_{-}\psi_{+}^{I}+{\Gamma_{(+)}}^{I}_{KL}\partial_{-}\phi^{K}\psi_{+}^{L})+\psi_{+}^{{I}}G_{I\bar{J}}(\partial_{-}\bar{\psi}_{+}^{\bar{J}}+{\Gamma_{(+)}}^{\bar{I}}_{\bar{K}\bar{L}}\partial_{-}\bar{\phi}^{\bar{K}}\bar{\psi}_{+}^{\bar{L}})\right)\nonumber\\
&&-\frac{i}{2}\left(\bar{\gamma}^{\bar{B}}_{-}n_{A\bar{B}}(\partial_{+}\gamma^{A}_{-}+A_{I \ \ C}^{\ A}\partial_{+}\phi^{I}\gamma^{C}_{-})+\gamma^{A}_{-}n_{A\bar{B}}(\partial_{+}\bar{\gamma}^{\bar{B}}_{-}+A_{\bar{I} \ \ \bar{C}}^{\ \bar{B}}\partial_{+}\bar{\phi}^{\bar{I}}\bar{\gamma}^{\bar{C}}_{-})\right.\nonumber\\
&& \ \ \ +\left.W_{\bar{I}AB}\partial_{+}\bar{\phi}^{\bar{I}}\gamma^{A}_{-}\gamma^{B}_{-}-W_{I\bar{A}\bar{B}}\partial_{+}\phi^{I}\bar{\gamma}^{\bar{A}}_{-}\bar{\gamma}^{\bar{B}}_{-}\right)+\ldots
\end{eqnarray}
where ${\Gamma_{(+)}}^{I}_{JK}=\Gamma^{I}_{JK}+G^{I\bar{L}}H_{\bar{L}JK}=G^{I\bar{L}}(\partial_{K}G_{J\bar{L}}+\partial_{J}B_{K\bar{L}}-\partial_{K}B_{J\bar{L}})$ and the $\ldots$ refer to four-fermi terms which we will not need for the moment.

By construction, this action is invariant under target space local Lorentz transformations,\footnote{For the local Lorentz transformations it is convenient to introduce a vielbein, $e^{a}_{I}$ and define $\psi^{a}_{+}=e^{a}_{I}\psi^{I}_{+}$.} target space gauge transformations, and target space Kalb-Ramond transformations:
\begin{eqnarray}\label{tgsym}
\psi^{a}_{+}&\rightarrow&\Lambda(\phi)^{a}_{\ b}\psi^{b}_{+}\nonumber\\
\gamma^{A}_{-}&\rightarrow&U(\phi)^{A}_{\ B}\gamma^{B}_{-}\nonumber\\
B&\rightarrow&B+d\omega_{1}.
\end{eqnarray} 
Note that plugging these transformations into the Lagrangian does not leave the action invariant -- rather, we must supplement these transformations on the worldsheet fields with a corresponding change in the background fields, $G$, $B$, $n$, $A$, and $W$. Only under this combined transformation is the Lagrangian, classically, invariant.

Quantum mechanically, however, this is not in general a symmetry of the theory due to an anomaly in the fermionic measure \cite{MooreNelson:1984,Moore:1984}.  Nonetheless,  these theories can be improved, order by order in $\alpha^{\prime}$ \cite{Hull:1985,Sen:1985,Hull:1986}, so as to respect a modified symmetry.  More precisely, while the theory is not invariant under (\ref{tgsym}), it is invariant under a slightly different symmetry whose transformation rule for $B$ is modified at order $\alpha^{\prime}$, 
\begin{eqnarray}
\delta B_{IJ}&\sim&\alpha^{\prime}\left(\partial_{[I}\theta^{M} A^{M}_{J]}-\partial_{[I}\theta^{ab}\omega^{ab}_{J]}\right).
\end{eqnarray}
where $\theta^{ab}$, $\theta^{M}$ parameterize infinitesimal local Lorentz transformations and gauge transformations respectively. With this transformation law, $H_{0}=dB$ is no longer gauge invariant. Correspondingly, we must also modify the definition of $H$ at order $\alpha^{\prime}$,
\begin{eqnarray}
H=H_{0}+\alpha^{\prime}\left(\Omega_{3}(\omega)-\Omega_{3}(A)\right)\,,
\end{eqnarray}
where $\Omega_{3}$ are the Chern-Simons three forms for the spin and gauge connections. The redefined H gives the modified Bianchi identity,
\begin{eqnarray}
dH=\alpha^{\prime}(tr R\wedge R-Tr F\wedge F).
\end{eqnarray}
This is the worldsheet manifestation of the Green-Schwarz mechanism \cite{Green:1984}. Note that if we wish to construct a theory with a non-anomalous symmetry that reduces to the classical transformations as $\alpha^{\prime}\rightarrow0$, these modifications are unavoidable. As emphasized in $\cite{Sen:1985}$ care must be used to see that supersymmetry is preserved.

In principle, then, there is no obstruction to studying \nK\ geometries with intrinsic torsion via a worldsheet NLSM -- the NLSM itself is perfectly well-posed, supersymmetric, and non-anomalous under all the symmetries of interest.  In practice, however, this construction is not very useful for many computational purposes.  First, as usual with a NLSM, most things we would like to compute end up depending on the physical metric; however, finding a solution to the Einstein equation and modified Bianchi is even more difficult than finding a compact Calabi-Yau metric (and indeed we still do not have a general proof of the existence of solutions to these equations except in a few very special cases \cite{Fu:2006}).  Second, since the Bianchi identity mixes orders in $\a'$, it is not clear when, or if, perturbation theory around a classical solution even makes sense.  Computationally, then, the NLSM is just not enough.

\subsection{Adding $dH$ to a \ZT\ GLSM by hand: the Green Schwarz mechanism}\label{byhand}

The obvious question, then, is how to implement $H$-flux and the modified transformation law for the $B$-field in a computationally effective GLSM.  Ideally, this would produce a GLSM which manifestly reduces to an NLSM of the above form at low energies.

To begin, consider a \ZT\ GLSM whose classical action reduces, at lower energies, to an effective NLSM for a complex manifold, $X$, with left-moving fermions valued in sections of a holomorphic vector bundle, $\CV_{X}$. Importantly, since all the geometry and topology of the IR CFT is generated by the gauge action, all of the potential anomalies are similarly embedded within the gauge group \cite{Manohar:1984}.  This allows us to lift the problem of tracking anomalies in the NLSM to the easier problem of identifying gauge anomalies in the UV-free GLSM.  This suggests that the anomalous Lorentz transformation law for the $B$-field in the NLSM should lift to an anomalous gauge transformation law for the $B$-field in the GLSM.

Our first job, then, is to locate the $B$-field in the GLSM.  Happily, this is well-understood physics: the $B$-field in such a GLSM is controlled  \cite{Witten:1993,Hori:2000}\ by the axial part of the FI coupling, 
$\CL_{\theta} =  \theta_{a}\, F^{~a}_{+-}$.  
The resulting spacetime $B$-field is then $B = \theta_{a} \omega^{a}$, where the $\w^{a}$ are the $(1,1)$-forms on the target space corresponding to the gauge field strengths, $F^{~a}_{+-}$.  More generally, the hermitian $(1,1)$ form (which becomes the complexified K\"ahler class on the $(2,2) $ locus) is $\CJ\equiv J+iB=t_{a}\w^{a}$, where $t_{a}=r_{a}+i\theta_{a}$ are the FI parameters in the superpotential.  Since $\w^{a}$ is closed and $\theta$ is constant, $H$=$dB$=0.

To introduce non-trivial $H$-flux, then, we can simply promote the $\theta_{a}$ to dynamical fields such that $H=dB=d\theta_{a}\wedge\omega^{a}$ does not vanish identically \cite{Hori:2001, Hori:2002,Adams:2006}.  Each such dynamical $\theta_{a}$ then represents a coordinate on an $S^{1}$ on the target space of the GLSM.\footnote{In the UV $\theta$ provides a new $S^{1}$ in the field space, in the IR this becomes an $S^{1}$ on the target space of the NLSM to which the theory flows, with the gluing in of the $S^{1}$ specified by the gauge action.}  Notably, this also generates the dilaton gradients we expect from supergravity, as shown \eg\ in \cite{Hori:2001}.  Unfortunately, gauge invariance and single-valuedness of the action require that $d\,(d\theta_{a})=0$,  and thus $dH=0$.

This suggests a simple way \cite{Adams:2006, AdamsLapan:2009, AdamsGuarrera:2009, Adams:2009} to build a GLSM with $dH\neq0$: let $\theta_{a}$ be shift-charged under the worldsheet gauge symmetry, $\theta_{a}\stackrel{\a}{\to}\theta_{a}+\mathcal{Q}_{a b}\,\a^{b}$.  
{
The resulting $B$ field now transforms non-trivially under the gauge symmetry, 
\be
B \,\stackrel{\a}{\longrightarrow}\, B+\mathcal{Q}_{a b}\,\a^{b}\w^{b}
\ee
}
The worldsheet gauge-invariant $H$-flux is thus of the form 
\be\label{E}
H = (d\theta_{a} +\CQ_{a b}\CA^{b})\wedge \w^{a}.
\ee 
Taking a further exterior derivative then gives
\be
dH = \CQ_{ab}\,{\w^{a}}\wedge\, \w^{b}\,,
\ee
where $\w^{b} = d\CA^{b}$ is the 2-form representing the gauge field strength under which $\theta$ is charged.

By making $\theta$ charged, however, we have rendered the classical action (in particular, the axial term $\CL_{\theta}$) non-gauge-invariant.  Fortunately, the variation of the axial term is precisely of the form of a 2d anomaly: under a gauge transformation with parameter $\a$, 
\be
\delta_{\a} \CL_{\theta} \propto \int \,{\CQ}_{ab}\,\a^{a}\,F^{~b}_{+-}\,.
\ee  
It is thus possible to cancel the gauge-variation of the classical action against a quantum anomaly in the measure {\em a la} Green and Schwarz 
\cite{Adams:2006}\footnote{Note that adding a GS term is not possible in a \TT\ model. Corespondingly, the fermionic spectrum is necessarily non-chiral, which forbids any gauge anomaly. Such models necessarily have $dH=0$.}, with $\CL_{\theta}$ providing the classical Green-Schwarz term.

This raises an obvious question: what is the gauge anomaly in a classical \ZT\ GLSM measuring?  It has long been understood that the gauge anomaly in a standard \ZT\ GLSM is in fact a probe of the sigma model anomaly in the target space, as follows.  The anomaly in 2d comes from a di-angle diagram and thus defines a quadratic form on gauge fields: $\CA=\CA_{ab}\,F_{+-}^{a}\,F_{+-}^{b}$.  Since the gauge fields represent the pullback to the worldsheet of 2-forms $\w^{a}$ in the target space, $F_{+-}^{a}=\phi^{*}\w^{a}$, the anomaly thus defines a 4-form on the target, $\CA = \CA_{ab}\w^{a}\w^{b}$.  A short computation (see \eg\ \cite{Adams:2006, Adams:2009}) then verifies that the corresponding 4-form is the RHS of the heterotic Bianchi identity, $\CA = \[{\rm tr} R\wedge R -{\rm Tr}F\wedge F\]$.\footnote{Technically, this result uses the natural \Ka\ structure on the toric variety.  In the presence of torsion, the physical metric will in general not be this \Ka\ metric.  They will differ, however, only by terms proportional to $\a'$, the loop counting parameter in the worldsheet; for the Bianchi identity above, we need only the leading order result.  Note that this argument is not reliable away from large radius -- however, away from large radius, the geometric picture is itself not reliable so we should focus instead on the quantum consistency of the gauge theory.} 
Choosing the gauge-transformation of our dynamical theta angle to cancel the worldsheet anomaly then ensures that we satisfy the target space Bianchi identity, with the vanishing of the net anomaly four-form in cohomology corresponding to an integrability condition for a smooth $H$.  Rather poetically, then, the Green-Schwarz mechanism on the target space pulls back to the GLSM as a Green-Schwarz mechanism on the worldsheet.  
This mechanism was first used in \cite{Adams:2006}\ to build GLSMs for \nK\ $T^{2}$-fibrations over $K3$ (which geometries were first studied via supergravity in \cite{Dasgupta:1999,Fu:2006,Goldstein:2002,Becker:2006}) and was subsequently exploited to study LG-orbifold points in the moduli space \cite{Adams:2009}, to compute the stringy spectrum \cite{AdamsLapan:2009},  and generalized to non-abelian GLSMs \cite{AdamsGuarrera:2009}.

\subsection{On the geometry of GS GLSMs}

We would like to argue that the Green-Schwarz mechanism for the worldsheet gauge theory, together with the anomalous transformation law for the $B$-field under worldsheet gauge variations, reduces precisely to the anomalous transformation law for the $B$-field and the corresponding Green-Schwarz mechanism of the NLSM.  To see that this indeed works out, we need to think more carefully about the effective geometry of the GS GLSM.

Consider an arbitrary, potentially anomalous, \ZT\ GLSM. At the classical level, this defines a classical NLSM by solving the equations of motion for all massive fields and evaluating the action on-shell. By construction, the target space $X$ of the resulting NLSM is simply the quotient of the flat target space of the UV  free fields (call it $Z$) by the (complexified) gauge group, $G$, of the GLSM, i.e. $X=Z^{\prime}/G$, where $Z^{\prime}$ is $Z$ with the fixed locus of G removed. If there is a superpotential turned on, we further restrict to the vanishing locus of the superpotential in $X$, $X|_{W=0}$.  The result is a completely garden-variety classical \ZT\ NLSM.

Quantum mechanically, of course, this NLSM may be anomalous via the standard sigma model anomaly.  The key observation \cite{Manohar:1984}\ is that, due to linearity of the UV theory, the sigma model anomaly of the NLSM is embedded in the gauge anomaly of the GLSM.  Moreover, from general properties of the classical geometry associated to a GLSM, we have a direct map between forms in the classical geometry, $X$, and gauge field strengths in the GLSM.  This allows us to translate the anomalous gauge-transformation law of the axial GS term in the GLSM directly to an anomalous transformation law of the $B$-field in the resulting classical NLSM.  It also allows us to relate the gauge-anomaly of the fermion measure in the GLSM to the sigma-model anomaly of the fermion measure in the NLSM.  The conditions for cancellation of the GLSM anomaly in the UV then directly map to the conditions required for cancellation of the sigma model anomaly in the IR NLSM as discussed above.

Of course, the above analysis only works at the classical level -- matching the two theories precisely would require showing that the full one-loop effective actions, including un-protected kinetic terms, precisely match up to finite renormalizations.  The corresponding calculation is straightforward in principle, but quite involved in practice, so we will not attempt it here.  Instead, we make the following observation.

Consider a classical NLSM constructed from the gauge-invariant on-shell action of a classical GLSM.  In general, both the NLSM and GLSM will have quantum anomalies.  As described above, we now have two ways of building non-anomalous models: we can cancel the gauge anomaly in the GLSM via a GS term for the worldsheet gauge symmetry; or we can cancel the sigma model anomaly of the NLSM directly using standard NLSM techniques as described above.  The question is whether these two quantum modifications are equivalent or not -- more precisely, whether they lie in the same universality class in the deep IR.  This can be represented in the following diagram:
\begin{eqnarray}\label{geodia}
\begin{ComD}
Classical \ GLSM @>\alpha^{\prime} \ Improve>> GLSM\\
@VVOn-ShellV @|\\
Classical \ NLSM @>\alpha^{\prime} \ Improve>> \ZT\  NLSM
\end{ComD}
\end{eqnarray}
While it remains technically possible for these models to be inequivalent, being different would mean there is a new way to deal with sigma model anomalies which is distinct from anything done before.  We consider this highly unlikely, and thus happily conjecture that the non-anomalous GLSM and NLSM thus defined in fact flow to the same CFT in the IR.

Finally, it is important to note that these GLSMs with non-trivial GS anomaly terms are not in fact linear.  When the mass of the heavy fermions is taken to infinity, however, corresponding to running to a boundary of the vector bundle moduli space, the only remaining non-linearity appears in the GS term itself, which will generically include globally ill-defined logs.  As we shall see in the next section, this need not cause us to panic.  However, various of the standard moves used in studying GLSMs must be considered with some care in this quasi-linear setting.  Happily, as we shall now explore in detail, we can in fact embed these quasi-linear models within completely standard \ZT\ GLSMs, with the quasi-linear models arising as low-energy effective descriptions of part of the moduli space.  The question thus reduces to asking whether the corresponding limits of the vector bundle moduli are well-behaved.

\subsection{Generating $dH$ in a garden-variety \ZT\ GLSM}

Surprisingly,  the GS couplings studied above, and the non-trivial $dH$ they represent, have been hiding in plain sight in almost all \ZT\ GLSMs studied to date.  
For example, consider a \ZT\ GLSM for a Calabi-Yau, $X$, at standard embedding, $\CV_{X}=T_{X}$.  Thanks to (\ref{EQ:BI}), standard embedding ensures  that $dH=0$.  
Crucially, such models generically contain vector bundle moduli  which deform the vector bundle away from standard embedding, $\CV_{X}=\tilde{T}_{X}$.
Away from standard embedding, however, we no longer have $tr R\wedge R = Tr F\wedge F$.  Consistency of the IR CFT, as reviewed above, then requires that  $dH\neq0$. But the CY GLSM we started with has no anomaly, and no GS terms, from which to derive a non-trivial $dH$.  Where is the $H$-flux hiding in this simple GLSM?

The key observation is that deforming the vector bundle boils down, in the GLSM, to tuning a set of Yukawa couplings which control the masses of a host of charged chiral fermions.  When we study the system at energies beneath the mass of some particular pair of fermions, it is appropriate to integrate them out.  However, if these heavy fermions transform in chiral representations of the gauge group (as for example does the top quark in the standard model), then integrating them out leaves us with an apparently anomalous spectrum of light fermions.  However as observed in \cite{D'HokerFarhi:1984} in the context of the standard model below the top mass, this would-be anomaly is precisely cancelled by a GS term generated when integrating out the heavy chiral fermions.  We are thus left with an effective GLSM in a Higgs phase with an anomalous spectrum of light fermions plus a GS term which ensures cancellation of the anomaly.  This is precisely the form of the models constructed by hand above.

It is useful to see this happen in detail in a simple example.  
Consider a garden-variety \ZT\ GLSM for a Calabi-Yau, $X$.  Since the specific model will not be important for what follows, we will be relatively schematic for the moment, focusing only on the details we will need.  For clarity of presentation, we restrict attention to a model with a single $U(1)$.  
The field content includes 
a set of chiral superfields, $\Phi_{I}$, a set of fermi fields, $\Gamma_{-A}$, and a $U(1)$ vector with fermi field strength $\Upsilon$.  Let the charges of the chiral and fermi fields be $Q_I$ and $q_A$ respectively (some of which may vanish), and let the fermi multiplets satisfy the chiral constraint, $\bar{D}_{+}\Gamma_{-A} = E_{A}\(\Phi_{I}\)$, as usual.
The corresponding Lagrangian has kinetic terms
\be
\mathcal{L}_{kin}=-\frac{1}{2}\int d^{2}\theta \left(
ie^{Q_{I}V}\bar{\Phi}_{I}\mathcal{D}_{-}\Phi_{I}
+e^{q_{A}V}\bar{\Gamma}_{-A}\Gamma_{-A}
+\frac{1}{4e^{2}}\bar{\Upsilon}\Upsilon
\) \,,
\ee
together with a chiral superpotential of the form,
\be
\CL_{W}=-\frac{1}{2}\int d\theta^{+}~\(
\mu \ \Gamma_{-A}J^{A}(\Phi)  +{1\over 4} t\Upsilon
\)   \ +h.c.
\ee
Here, the $J^{A}(\Phi)$ are holomorphic functions of the $\Phi_{I}$ with net gauge charge $-q_A$ such that the superpotential is gauge-invariant.  Chirality of the superpotential then requires that $E_{A}J^{A}=0$.
The conventions and component expansions are listed in Appendix (\ref{Sec:Conventions}).\footnote{A useful tool for performing \ZT\ component expansions in mathematica with examples may be found at \url{http://www.mit.edu/~edyer/code.html}.} 

Since the fermions in such a \ZT\ gauge theory live in chiral representations of the gauge group, there will in general be a gauge anomaly.  Under a general gauge transformation with gauge parameter $\L$, the effect of the anomaly is to shift the action by,
\be
\delta \CL \propto \int d\theta^{+} \hat{\CA}\,\L\Upsilon~+h.c.\,,
~~~~~~~~
\hat{\CA}=\sum_{I}Q_IQ_I - \sum_{A}q_Aq_A\,,
\ee
as can be verified by a standard one-loop calculation.  As discussed above, we are interested in UV GLSMs which are completely free of anomalies, so we hereby demand that $\hat{A}=0$, as is the case, for example, at standard embedding.

Among the  terms that appear in the action are a host of Yukawa couplings,
\be
\int  d\theta^{+}  ~\m\, \Phi_{H}\Phi_{q}\G_{Q-q} ~+h.c. \,,
\ee
where $\Phi_{q}$ is a chiral field of charge $q$, $\G_{Q-q}$ is a fermi field of charge $Q-q$,  $\Phi_{H}$ is a chiral field of charge $-Q$ which gets a non-zero vev $\vev{\phi_{H}}$ on the chosen patch of the target space, and $\m$ is a tunable modulus.  Such terms can arise, for example, from $\Gamma_{A} {J}^{A}({\Phi})$ {terms in the action.}\footnote{Similar (non-superpotential) terms arise from or $E_{A}\bar{\Gamma}^{A}_{-}$ terms deriving from the kinetic terms for the fermi multiplets; the basic effect is directly analogous (though details can be phenomenologically different), so we focus for simplicity on the superpotential case for the moment.}   
Below the scale set by $\m\vev{\phi_{H}}$, we can integrate out $\Phi_{q}$ and ${\G}_{Q-q}$ to get an effective action for the remaining light modes.  

For example, consider a \ZT\ model for a Calabi-Yau at standard embedding, corresponding to a \TT\ point in the moduli space.  At such points, the mass terms for the fermions are non-chiral,
\begin{eqnarray}
\CL_{yuk} = \sqrt{2}\int d\theta^{+} \left(\partial_{I}G \ \Gamma_{-}^{I}P+\Pi_{-}G\right)+h.c.&\propto&  \partial_{I}G \ \( \pi_{-} \psi_{+}^{I} + \  \psi_{-}^{I}  \pi_{+} \) + \ldots
\end{eqnarray}
where the $\pi_{\pm}$ have charge $-Q$, the $\psi_{\pm}$ have charge 1, and $G=0$ defines a hypersurface in X.
Since the scalar coefficient is the same for both pairs of fermions, the masses are the same, so while integrating out either pair of unequally charged fermions would generate an anomaly, the two pairs are degenerate so there is no regime in which it makes sense to integrate out one but leave the other.

On the other hand, if we turn on a deformation which breaks this accidental \TT\ supersymmetry, these two chiral mass terms will no longer be degenerate. A simple such $(0,2)$ deformation, for example, replaces $\p_{I}G$ in the $\G_{-}^{I}$ term with a general function, $(\p_{I}G+\m\, J_{I})$. 
The coupling $\m$ represents a modulus of the vector bundle, with the precise geometry of this deformation being encoded in the functional form of $J_{I}$.  
Adding this deformation into the action then gives,
\begin{eqnarray}
\mathcal{L}_{yuk}&\propto&  \partial_{I}G \( \pi_{-} \psi_{+}^{I}\) + \(\p_{I}G+\m\, J_{I}\) \( \psi_{-}^{I}  \pi_{+} \) + \ldots
\end{eqnarray}
When $\m=0$, we recover $(2,2)$ supersymmetry and a non-chiral mass spectrum.  When $\mu$ is non-zero, on the other hand, the masses of the two pairs of chiral fermions will be different.  When $\m$ is large we will generally find that one pair is heavy and should be integrated out while the other remains light and fluctuates.  At energies well below the mass of the heavy fermions, the spectrum of surviving light fermions is thus explicitly anomalous.  

As the heavy fermions are gauge-charged, integrating them out requires some care.  One method is to use the non-vanishing Higgs field to change variables to uncharged fields,

\be
\Phi_{q} = (\Phi_{H})^{q\over Q} \, \tilde{\Phi}\,,
~~~~~~~~
{\G}_{Q-q} = (\Phi_{H})^{Q-q \over Q}\, \tilde{\G}\,,
\ee
which we can then integrate out without concern.  However, in doing this change of variables we pick up a Jacobian factor of the form,
\be
\CD\Phi_{q} \,\CD\G_{Q-q} 
=
\CD\tilde{\Phi} \,\CD\tilde{\G}  \, e^{\,\hat{C}\!\int \log(\Phi_{H})\,\Upsilon}
\ee 
where $\hat{C}$ is such that the gauge variation of the resulting GS term precisely cancels the effective action of the remaining light fermions.  (See \cite{D'Hoker:1984, D'HokerFarhi:1984} for a lovely example of this effect from integrating out the heavy top quark in the standard model.)  These anomaly-canceling GS terms then descend, in the deep IR, to non-trivial $dH$ satisfying the Bianchi identity, as discussed in Sec. 2 and in  \cite{Adams:2006,Adams:2009}.

Crucially, for the special case of standard embedding (or indeed at any \TT\ point in the \ZT\ moduli space), the Yukawa couplings are tightly constrained such that the {physical} fermion masses and interactions are non-chiral.  There is thus never any regime of parameter space or energy in which the set of fermions above, or below, a given mass is chiral.  When integrating out the massive fermions, then, all such $H$-flux generating terms must cancel.

Note that the resulting effective model has an interesting limit in which we send our bundle parameter to infinity, $\mu \to\infty$.  In this limit, the massive fermions entirely decouple and we are left with precisely the theory we started with, minus the fields $\Phi_{q}$ and $\G_{Q-q}$, but plus the GS term $\CL_{GS}=\hat{C}\log(\Phi_{H})\Upsilon$.  But this is precisely the form of the theories studied in \cite{Silverstein:1995,Blaszczyk:2011, Quigley:2011}: the anomaly arising from the measure for a chiral family of fermions is cancelled by a logarithmic Green-Schwarz term, where the argument of the log is a function of the charged scalar fields in the theory.  It should now be clear that virtually all such models can be orchestrated by Higgsing and integrating out a chiral set of fermions, as above.  It is also clear what physics lies behind the singularities at points where the logarithms diverge: such singularities signal the return of the again-massless fermions we had previously integrated out: death from the UV.

Note, too, that the resulting models should be treated with some caution.  In particular, the \ZT\ GLSM with which we began is as well-behaved as one could hope. Fore example, it enjoys a non-singular topological chiral ring which varies smoothly with the moduli, and the unbroken (0,2) supersymmetry together with the linear model structure ensre that world sheet instantons do not generate a spacetime superpotential \cite{Basu:2003,Beasley:2003}. However, once we take the limit $\mu\to\infty$, the assumptions going into those arguments no longer trivially hold, so these results must be re-evaluated.  Doing so, however, is hard: since the GS terms are non-linear, and indeed logarithmic, computing the quantum OPEs does not trivialize in the UV.  This motivates us to search for more tractable variants of these models which have more gentle, and computable, physics in the UV.

%
%
%

\section{Verifying Quantum Consistency in a Special Class of Models}\label{Sec:Models}
 
As we have seen, moving to a generic point in the $(0,2)$ moduli space leaves us with a \nK\ manifold supporting non-trivial $H$-flux.  Around such points, it is sometimes convenient to integrate out the heavy fermions, generating a quasi-linear Green-Schwarz model in which the structure of the $H$-flux is more manifest.  The price of this simplicity is losing the manifest good-behavior of the original GLSM.  It is thus illuminating to verify that these effective GS models are in fact good quantum-consistent GLSMs which flow to candidate CFTs in the IR.  

In this section we marshall evidence that these GS effective theories do, in fact, flow to good quantum-consistent CFTs in the IR.  For simplicity, we focus on a special class of such GS models which are fully linear (avoiding the logarithms which generically appear in such effective models).  Borrowing a technique from Silverstein and Witten \cite{Silverstein:1994}, and following a similar analysis to that in \cite{Adams:2009,Hori:2001}, we identify a chiral left-moving conformal algebra in the UV that has all the properties needed to flow to a left-moving conformal algebra in the IR. The existence of this algebra is equivalent to the vanishing of anomalies for specific left moving and right moving symmetries. We briefly review the justification of this technique. We then present the details of the calculation in the case of a single $U(1)$ gauge group, then generalize the results to the case of multiple $U(1)$s. We also compute and record the left and right central charges, and vector bundle rank, of the IR fixed point theory as a function of the charges in the original quasi-linear model.  We begin by identifying the linear models of interest.

\subsection{The Models}

Consider the following models, which will be the focus of the rest of this section.  As above, we begin with chiral superfields, $\Phi_{I}$, fermi fields, $\Gamma_{-A}$, and a set of $U(1)$ vectors with fermi field strengths, $\Upsilon_{a}$, under which the matter fields have charges $Q^{a}_{I}$ and $q^{a}_{A}$.  The fermi multiplets again satisfy the chiral constraint, $\bar{D}_{+}\Gamma_{-A} = E_{A}\(\Phi_{I}\)$.  
The Lagrangian includes kinetic terms,
\be
\mathcal{L}_{kin}=-\frac{1}{2}\int d^{2}\theta \left(
ie^{Q_{I}^{a}V_{a}}\bar{\Phi}_{I}\mathcal{D}_{-}\Phi_{I}
+e^{q_{A}^{a}V_{a}}\bar{\Gamma}_{-A}\Gamma_{-A}
+\frac{1}{4e^{2}_{a}}\bar{\Upsilon}_{a}\Upsilon_{a}
\) \,,
\ee
where the $e_a$ are the gauge couplings, together with a chiral superpotential of the form,
\be
\CL_{W}=-\frac{1}{2}\int d\theta^{+}~\(
\mu \ \Gamma_{-A}J^{A}(\Phi)  +{1\over 4} t^{a}\Upsilon_{a} 
\)  +h.c.
\ee
Here, the $J^{A}(\Phi)$ are holomorphic functions of the $\Phi_{I}$ with net gauge charge $-q_A$ such that the superpotential is gauge-invariant.  Chirality of the superpotential then requires that $E_{A}J^{A}=0$.
The conventions and component expansions are listed in Appendix (\ref{Sec:Conventions}). 

Since the fermions live in chiral representations, the gauge symmetry will again be anomalous with anomaly matrix $\hat{\CA}^{ab}=Q^{a}_{I}Q^{b}_{I} - q^{a}_{A}q^{b}_{A}$.  
Instead of setting the anomaly to zero, however, we now add to the model a set of Green-Schwarz axions, $Y_{\ell}$, with shift-charges $\mathcal{Q}^{a}_{\ell}$,
\begin{eqnarray}
\delta_{\Lambda} Y_{l}&=&-i\mathcal{Q}_{l}^{a}\Lambda_{a}\\
\mathcal{D}_{-}Y_{l}&=&\partial_{-}Y_{l}+\frac{\mathcal{Q}_{l}^{a}}{2}(\partial_{-}V_{a}+iA_{a})\,,
\end{eqnarray}
standard gauge-invariant kinetic terms,
\be
\mathcal{L}_{Y}=-\frac{i}{4}\int d^{2}\theta^{+} \, {k^{2}_{l}}(Y_{l}+\bar{Y}_{l}+\mathcal{Q}_{l}^{a}V_{a})\mathcal{D}_{-}(Y_{l}-\bar{Y}_{l})
\ee
and a set of non-gauge-invariant Green-Schwarz terms,
\be
\mathcal{L}_{GS}=\frac{1}{4}\int d^{2}\theta^{+}\hat{\mathcal{C}}^{[a}_{l}\mathcal{Q}^{b]l}V_{a}A_{b}-\frac{1}{4}\hat{\CC}^{al} \int d\theta^{+}\Upsilon_{a} Y_{l}+h.c.\\\ \label{SFLagend}
\ee
The $\hat{\mathcal{C}}^{a}_{\ell}$ specify the axial couplings of the axions to the gauge fields.  Note that the anomaly itself is strictly symmetric, while the axial term in $\CL_{GS}$ has a priori no symmetry.  The purpose of the $VA$ term in $\CL_{GS}$ is to cancel the antisymmetric part of the axial term, leaving the symmetric part to cancel the quantum anomaly.

The $Y_{\ell}$ fields may be thought of as St\"uckelberg fields for the anomalous gauge multiplets. In particular, for any anomalous $U(1)$ we can introduce a shift charged field, $Y$, together with suitable GS couplings, so that the anomaly cancels the variation of the classical action. Note that setting the gauge $Y=0$ we return to the original anomalous theory, though now with an explicit mass term arising as the legacy of $Y$'s kinetic term. For more details about quantizing anomalous theories and applying the St\"uckelberg method to an anomalous action in the non-supersymmetric case see \cite{Preskill:1990}.

\subsubsection{Example: A Single $U(1)$}
To gain a little familiarity with these shift fields, let's take a brief look at the classical geometry in a simple gauge invariant case, a single U(1) and shift field of charge, $\mathcal{Q}=1$, but no GS coupling, and then highlight the subtlety when including the anomaly. The action and component expansions are given in Appendix (\ref{actions}). For this simple model, the bosonic potential is given by:
\bea
U = \frac{e^2}{8}\(\sum_{I}Q_{I}|\phi_{I}|^{2}+k^{2}(y+\bar{y} ) - r\)^2 + \frac{\mu^{2}}{2}\sum_A |J_A|^2  + \sum_A |E_A|^2.
\eea
The classical moduli space is obtained by restricting the field configuration to minimize the bosonic potential: $D=0$, $J_A=0$, and $E_A=0$. In focusing on the D-term constraint, we see that there are no compact models with a single $U(1)$ and a shift field. There always exists a runaway direction. The vanishing of  $J_A$ and $E_A$ cannot help with compactness, as the holomorphic hyper-surface of a non-compact complex manifold is either a discrete set of points or non-compact.

Now consider the same model as above, but with an anomalous fermion content cancelled by a GS term for $Y$,
\begin{eqnarray}
\mathcal{L}_{GS}&=&-\frac{1}{4}\hat{\mathcal{C}}\int d\theta^{+}\Upsilon Y + h.c.
\end{eqnarray}
One might naively be tempted to plug in the Wess-Zumino (WZ) component expansion for $\Upsilon$ and $Y$. 
\begin{eqnarray}
\mathcal{L}_{GS}&=&-\frac{\hat{\mathcal{C}}}{2}\left(D(y+\bar{y})-iF_{+-}(y-\bar{y})+i\lambda_{-}\chi_{+}+i\bar{\lambda}_{-}\bar{\chi}_{+}\right)\nonumber\\
\end{eqnarray}
Yielding,
\begin{eqnarray}
D&=&-\frac{e^{2}}{2}(\sum_{I}Q_{I}|\phi_{I}|^{2}+(k^{2}- \hat{\mathcal{C}})(y+\bar{y})- r ).\nonumber
\end{eqnarray}
This doesn't make sense classically, however. The Lagrangian is not gauge invariant, and so fixing WZ gauge in the classical action is not possible.

It is of course possible to do away with gauge symmetry completely and write down the full component expansion. After appropriate field redefinitions, the component action is:
\begin{eqnarray}
\mathcal{L}_{GS}&=&-\frac{\hat{\mathcal{C}}}{2}\left(D(y+\bar{y})-iF_{+-}(y-\bar{y})+i\lambda_{-}\chi_{+}+i\bar{\lambda}_{-}\bar{\chi}_{+}\right)\nonumber \\
&+&\frac{\hat{\mathcal{C}}}{2}\left(D(s+\bar{s})-iF_{+-}(s-\bar{s})+i\lambda_{-}\zeta_{+}+i\bar{\lambda}_{-}\bar{\zeta}_{+}\right).
\end{eqnarray}
Where $s$ and $\zeta_{+}$ encode the unfixed parts of the gauge field. This is going too far, however. This action, and the corresponding modified D term,
\begin{eqnarray}
D&=&-\frac{e^{2}}{2}(\sum_{I}Q_{I}|\phi_{I}|^{2}+(k^{2}- \hat{\mathcal{C}})(y+\bar{y})+\hat{\mathcal{C}}(s+\bar{s})- r ),
\end{eqnarray}
should not be thought of as a good starting place for analyzing the topology and geometry. In writing these classical expressions down we have neglected crucial one loop effects, not least of which is the fact that the full theory is gauge invariant. 

\subsubsection{Y multiplets vs $T^2$  multiplets}

In the models introduced in this section, the Green-Schwarz axion field is a \ZT\ chiral boson with a shift gauge symmetry: 
\be
Y = y + \sqrt{2} \tp \chi_+ - i \tp \tbp \p_+ y, \quad e^{y}   \in \mathbb{C}^*.
\ee
This is a particular supersymmetric completion of the GS axion. Instead, we could have chosen a different completion. For example, the torsion multiplet, $\Theta$ in  \cite{Adams:2006, Adams:2009, AdamsLapan:2009} of the $T^2$ models is another completion,
\bea
\Theta = \vartheta + \sqrt{2}\tp \tilde \chi_+ - i \tp \tbp \p_+ \vartheta, \quad \vartheta= \theta_1+  i \theta_2 \in T^2.
\eea
%
Choosing a different supersymmetric completion of the GS axion has a natural interpretation in terms of quotient actions of the target space. $\CN=(0,2)$ ensures that the target space is a complex manifold; precisely what the complex structure is follows from the action of SUSY on the real scalars; fixing this SUSY action then fixes the action of the complexified gauge group, and thus determines the topology of the quotient. Inequivalent SUSY completions of the GS-axion thus correspond to different quotient actions.  The effective geometry of the $T^2$ multiplet is analyzed carefully in  \cite{Adams:2006}. 

\subsection{Methodology}
The algebra of a $(0,2)$ superconformal field theory consists of a left moving stress tensor, $\mathcal{T}_{L}$, a right moving stress tensor, $\mathcal{T}_{R}$, right moving supercurrents, $\mathcal{G}_{R}^{\pm}$, and a right moving R current $\mathcal{J}_{R}$. We will rely on the existence of an additional left moving current, $\mathcal{J}_{L}$, to have some control in the IR. These operators satisfy the following OPE algebra:

\begin{eqnarray}
\begin{array}{l c l l c l}
\mathcal{J}_{L}(x^{-})\mathcal{J}_{L}(y^{-}) & \sim & \frac{r_{L}}{(x^{-}-y^{-})^{2}} & \mathcal{J}_{R}(x^{+})\mathcal{J}_{R}(y^{+}) & \sim & \frac{1}{2}\frac{\hat{c}_{R}}{(x^{+}-y^{+})^{2}}\\
\\
 & & & \mathcal{J}_{R}(x^{+})\mathcal{G}_{R}^{\pm}(y^{+}) & \sim & \pm\frac{\mathcal{G}_{R}^{\pm}(y^{+})}{x^{+}-y^{+}}\\
 \\
  & & & \mathcal{G}_{R}^{+}(x^{+})\mathcal{G}_{R}^{-}(y^{+}) & \sim & \frac{1}{2}\frac{\hat{c}_{R}}{(x^{+}-y^{+})^{3}}+\frac{\mathcal{J}_{R}(y^{+})}{(x^{+}-y^{+})^{2}}\\
   & & & & & +\frac{\mathcal{T}_{R}(y^{+})+\frac{1}{2}\partial_{+}\mathcal{J}_{R}(y^{+})}{x^{+}-y^{+}}\\
\\
 \mathcal{T}_{L}(x^{-})\mathcal{T}_{L}(y^{-}) & \sim & \frac{1}{2}\frac{c_{L}}{(x^{-}-y^{-})^{4}}+2\frac{\mathcal{T}_{L}(y^{-})}{(x^{-}-y^{-})^{2}} & \mathcal{T}_{R}(x^{+})\mathcal{T}_{R}(y^{+}) & \sim & \frac{3}{4}\frac{\hat{c}_{R}}{(x^{+}-y^{+})^{4}}+2\frac{\mathcal{T}_{R}(y^{+})}{(x^{+}-y^{+})^{2}}\\
 & & +\frac{\partial_{-}\mathcal{T}_{L}(y^{-})}{x^{-}-y^{-}} & & & +\frac{\partial_{+}\mathcal{T}_{R}(y^{+})}{x^{+}-y^{+}}\\
 \\
 \mathcal{T}_{L}(x^{-})\mathcal{J}_{L}(y^{-}) & \sim & \frac{\mathcal{J}_{L}(y^{-})}{(x^{-}-y^{-})^{2}}+\frac{\partial_{-}\mathcal{J}_{L}(y^{-})}{x^{-}-y^{-}} & \mathcal{T}_{R}(x^{+})\mathcal{J}_{R}(y^{+}) & \sim & \frac{\mathcal{J}_{R}(y^{+})}{(x^{+}-y^{+})^{2}}+\frac{\partial_{+}\mathcal{J}_{R}(y^{+})}{x^{+}-y^{+}}
\end{array}
\end{eqnarray}
In this section we identify theories which are believed to flow to such superconformal field theories with nontrivial central charge and vector bundle rank at their IR fixed point. We find these theories by constructing models that posses such an algebra in cohomology even in the UV. Though this is neither necessary nor sufficient to guarantee the existence of the IR algebra\footnote{It is conceivable that the UV \ZT\ algebra does not describe the IR fixed point; further operators may appear near the fixed point, the theory may become trivial, etc\ldots.}, it is usually taken as strong motivation\cite{Silverstein:1994,AdamsLapan:2009,Hori:2001, AdamsGuarrera:2009}.


The $(0,2)$ supersymmetry algebra contains the anti-commutation relation:
\begin{eqnarray}
\{\bar{Q}_{+},Q_{+}\}&=&2P_{+}.
\end{eqnarray}
As such, elements of $\bar{Q}_{+}$ cohomology are in one to one correspondence with left moving ground states. Furthermore, correlators of cohomology elements are protected.\footnote{Imagine that the action depends on some parameter, $t$, multiplying a $\bar{Q}_{+}$ commutator, $\mathcal{L}=\mathcal{L}_{0}+t[\bar{Q}_{+},\mathcal{O}]$, then correlators of $\bar{Q}_{+}$ cohomology elements are independent of $t$. A specific realization of this will allow us to compute the OPEs.} This motivates searching for the left moving components of the superconformal algebra in cohomology. It turns out that it is more convenient to identify states in $\bar{D}_{+}$ cohomology rather than $Q_{+}$ cohomology. This is not a problem, as the two operators are conjugate.


Even after finding candidate chiral currents, one might imagine that calculating the OPEs would be difficult in the presence of a superpotential and gauge interactions. As it turns out this is not an issue. Due to the magic of supersymmetry, there is a dramatic simplification when considering left moving ground states. As explained in \cite{Silverstein:1994}, the superpotential comes with a dimension-full parameter, $\mu$. By power counting, any term in the operator algebra that contains a factor of $\mu$ must also contain a factor of $x^{2}$ and so vanishes in the $x^{+}\rightarrow0$ limit, while the gauge interactions flow away in the UV. This means that we can use the free field OPEs in calculating the algebra.

So far this has just been a discussion of the left moving part of the algebra. As we will see in the following sections, however, the existence of this left moving algebra relies heavily on having a non anomalous R-symmetry. Once this R-symmetry is discovered, the rest of the right moving algebra is guaranteed, as long as supersymmetry is preserved. In the next few sections we walk through the construction of the currents and the calculation of their algebra for the models of interest.



\subsection{Gauge Invariant Model}
To begin with, let's examine the case of a single $U(1)$ and shift field coupled in a gauge invariant fashion. We consider a model with fairly generic field content:
\bea
\Phi_I, \qquad \Gamma_{-A}, \qquad Y 
\eea
with the usual chirality constraints 
\bea
\DBp\Phi_I = \DBp Y = 0, \qquad \DBp \Gamma_{-A} = \sqrt{2} \; E_A.
\eea
For now, we will focus on a single $U(1)$ gauge field. The action is
\bea\label{EQ:ActionNoGS}
 \CL &=&-\frac{1}{2}\int d^{2}\theta^{+}\left(i\sum_{I}e^{Q_{I}V}\bar{\Phi}_{I}\mathcal{D}_{-}\Phi_{I}+i\frac{k^{2}}{2}(Y+\bar{Y}+V)\mathcal{D}_{-}(Y-\bar{Y})+\sum_{A}e^{q_{A}V}\bar{\Gamma}_{-A}\Gamma_{-A} + \frac{1}{4e^2} \bar \U \U \right)\nonumber\\
 && - \frac{\mu}{2}\int d\theta^{+} \sum_{A}\Gamma_{-A}J^{A}+h.c. 
\eea
Here $\sum_{A}E_{A}J^{A}=0$ in order to preserve supersymmetry.

This action is classically gauge invariant. In what follows we will write down the conditions for the gauge symmetry to be anomaly free.
\subsubsection{Equations of Motion}
The equations of motion for these models are:
\bea
\DBp \(e^{q_A V}  \bar{\Gamma}_{-A} \) &=& -\mu \  J^A \\
\DBp \(e^{Q_I V} \Dm \bar \Phi_I \) &=& - i \mu \sum_A \Gamma_{-A} \p_I J^A + i \sqrt{2} \sum_A e^{q_A V} \bar{\Gamma}_{-A} \p_I E_A  \\
k^{2}\DBp \(\Dm \bar Y \) &=&   -i\mu \ \Gamma_{-A}\partial_{Y}J^{A} + i\sqrt{2}\sum_A e^{q_A V}\bar{\Gamma}_{-A} \partial_{Y}E_A \\
\f{1}{2e^2}\DBp \(\p_- \bar \U \) &=&   \sum _I e^{Q_I V} Q_I \Phi_I \Dm \bar \Phi_I+i \sum_A e^{q_A V} q_A \bar{\Gamma}_{-A} \Gamma_{-A} +k^2 \Dm \bar Y.
\eea
\subsubsection{Global Symmetries}
We are interested in models which posses global symmetries, $\Phi_{I}\rightarrow e^{-i\alpha_{I}\epsilon}\Phi_{I}$, in addition to the gauge symmetry. In particular we want candidate $U(1)_{L}$ and $U(1)_{R}$ symmetries.
We need further constraints on $E_A$ and $J^A$ for our theory to be invariant under these global symmetries. First of all, $\Gamma_{-A}$ and $J^A$ must have opposite charges; $J^A \(Y-i\kappa \epsilon, e^{-i \alpha_I \epsilon} \Phi_I\) = e^{ i \beta_A \epsilon}J^A$; while $E_A$ and $\Gamma_{-A}$ must have the same charge; $E_A\(Y-i\kappa \epsilon, e^{-i \alpha_I \epsilon} \Phi_I\) = e^{- i \beta_A \epsilon}E_A$. These relations imply the following  \emph{quasi-homogeneity conditions}. 
\bea
\sum_I \alpha_I \Phi_I \p_I J^A + \kappa\partial_{Y}J^{A} + \beta_A J^A = 0 \\
 \sum_I \alpha_I \Phi_I \p_I E_A + \kappa\partial_{Y}E_A -\beta_A E_A = 0 
\eea

Where $\kappa$ is the charge of the shift field, $Y$,  $\alpha_I$ are charges for the chiral scalars, $\Phi_I$, and $\beta_A$ are charges for the fermi multiplets, $\Gamma_{-A}$.

One needs a little more care for the R-symmetry due to the fact that both $\theta ^+$ and  $\DBp$ have R charge +1.  Preserving R-symmetry requires $\Gamma_{-A} J^A$ to have  R charge +1. Thus the $J^A$ have R charges $-  \beta^R_A +1$, while the $E_A$ have R charges $ \beta^R_A + 1$. The quasi-homogeneity conditions corresponding to the R-symmetry are:
\bea
\sum_I  \alpha^R_I \Phi_I \p_I J^A +  \kappa^R\partial_{Y}J^{A} + (\beta^R_A -1)J^A = 0 \nonumber \\
\sum_I \tilde \alpha_I \Phi_I \p_I E_A +  \kappa^R\partial_{Y}E_A - ( \beta^R_A+1) E_A = 0.  \nonumber 
\eea

Now that we have our equations of motion and quasi-homogeneity conditions we can begin to search for the chiral superconformal algebra. In particular we will identify chiral superfield currents $J_{L}^{+}$ and $T^{++}$, whose lowest components are gauge invariant conserved currents which satisfy the conformal OPE relations. 

We begin by discussing the various $U(1)$ currents in our theory,

\subsubsection{$U(1)$ Currents}

In our class of models, there are three $U(1)$ symmetries of particular interest:  $U(1)_G$ gauge,  $U(1)_R$ and global $U(1)_L$ symmetry. Each of these plays a critical role in constructing the IR theory and the $(0,2)$ superconformal algebra. The gauge symmetry effects a quotient of the target space, and ensuring gauge invariance at the quantum level is crucial. The $U(1)_{L}$ symmetry descends to the IR and can be used for the GSO projection in string backgrounds. As we will see, having a non-anomalous $U(1)_{R}$ current is essential for constructing the chiral, left moving stress tensor.

Let us start by exploring the gauge current.
\bea
j_G^{+}  &=& - i \sum_I  Q_I  \phi_I \Dm \bar \phi_I  - \sum_A q_A  \gamma_{- A} \bar \gamma_{- A } -  i k^2 \, \Dm  \bar y  \\
j_G^{-} &=&   i \sum_I Q_I  \bar \phi_I {\cal D}_+  \phi_I  - \sum_I  Q_I \psi_{+ I} \bar \psi_{+ I } + i k^2\, {\cal D}_+ y
\eea
Note, the current is neutral, so it is conserved both partially and covariantly. The superfield completion is
\bea
J_G^+ &=&- i \(\sum_I e^{Q_I V} Q_I \Phi_I \Dm \bar \Phi_I  + i \sum_A e^{q_A V} q_A \bar{\Gamma}_{-A} \Gamma_{-A} + k^2 \, \Dm \bar Y  \)   \\
J_G^-  &=& \f 12 \DBp\(\sum_I Q_I \bar \Phi_I \Dp \(e^{Q_I  V} \Phi_I\)+ k^2 \, \Dp \(Y+ V\) \) .
\eea

It is nice to see that $J_G^-$ is $\DBp$ trivial. Thus the interesting contribution of the gauge current in cohomology comes purely from $J_G^+$. 

 In order to check the chirality of $J_G^+$,  recall the EOM:
\bea
\f{1}{2e^2}\DBp \(\p_- \bar \U \) &=&   \sum _I e^{Q_I V} Q_I \Phi_I \Dm \bar \Phi_I+i \sum_A e^{q_A V} q_A \GB_ {-A} \Gamma_{-A} +k^2  \Dm \bar Y =  i J_G^+,
\eea
$J_{G}^{+}$ is exact up to equations of motion. $\DBp J_G^+ = 0$ follows from the fact that $\DBp^2 = 0 $.  Another way to see this is to apply $\DBp$ directly to $J_G^+$. Using the classical EOM and the quasi-homogeneity conditions, one can explicitly check that $J_G^+$ is chiral. 

For any other global symmetry, in particular for $U(1)_L$, the procedure to find a gauge invariant chiral current is similar and we get:
\bea
J_L^+ = -i \(\sum_I e^{Q_I V} \alpha^L_I \Phi_I \Dm \bar \Phi_I +i \sum_A e^{q_A V} \beta^L_A \bar{\Gamma}_{-A} \Gamma_{-A} + k^2 \kappa^L \Dm \bar Y \).
\eea
Classical chirality, $\DBp J_L^+|_{EOMs} = 0 $, can be checked using the EOM and the quasi-homogeneity conditions.

The R-current is  a little trickier, as the component fields come with different R charges. The lowest component of the R-current is: 
\bea
i j_R^+ &=& \sum_I \alpha^R_I \phi_I \mathcal{D}_{-}\bar \phi_I - i \sum_A \beta^R_A  \gamma_{- A} \bar \gamma_{-  A} + k^2 \kappa^R \mathcal{D}_{-}\bar y + \frac{i}{2e^2} \bar \lambda_- \lambda_- \\
ij_R^- &=& -\sum_I \(\alpha_I^R \phi_I \mathcal{D}_{+} \bar \phi_I - i \(\alpha_I^R-1\)\bar \psi_{+ I } \psi_{+ I }\) - k^2 \kappa^R \mathcal{D}_{+} y  - ik^2 \bar \chi_+ \chi_+
\eea
which is gauge invariant and conserved. 

\subsubsection{Further Modification of Chiral Currents }\label{sec:mod currents}
A classically chiral current for a given symmetry, $J_S^+$, can fail to be chiral in a quantum theory. The supersymmetric extension of the chiral anomaly, known as the Konishi anomaly\cite{Konishi:1983}, in 1+1 dimensions \footnote{ See, for example Appendix C of \cite{Hori:2001} and section 3 of \cite{Basu:2003}.}, tells us that
\bea
\langle \DBp J^+ \rangle = \langle  \frac{ \CA _S \U}{8 \pi} \rangle.
\eea
where $\CA _S$ is the anomaly coefficient for the given symmetry. 

Generically it is not possible to remove this anomaly by redefining the current in a manner that preserves gauge invariance. As noted in \cite{Hori:2001}, however, when there is a shift charged field in the game we can remove the anomaly.  This freedom to redefine the current is easy to understand by looking at the bosonic components. In two dimensions the anomaly is given by $\partial_{\mu}j^{\mu}\propto\epsilon^{\mu\nu}\partial_{\mu}A_{\nu}$. It is possible to define a new conserved current, $\tilde{j}^{\mu}=j^{\mu}-\epsilon^{\mu\nu}A_{\nu}$, but this is not gauge invariant. If we have a shift field, $\theta$, at our disposal we can do better and define, $\tilde{j}^{\mu}=j^{\mu}-\epsilon^{\mu\nu}(\partial_{\nu}\theta+A_{\nu})$. This is gauge invariant and conserved. This new current corresponds to adding a term of the form $\theta\epsilon^{\mu\nu}F_{\mu\nu}$ to the quantum effective action and then improving the resulting current. Note, that this addition to the action is just the form of a Green-Schwarz term.

The superfield completion of this effect is:
\bea
\tilde J_S^+ =  J_S^+ + i \f{\CN_S}{4\pi} \Dm Y.
\eea 
The modified currents is no longer (classically) chiral:
\bea
\DBp \(J_S^+ + i \f{\CN_S}{4\pi} \Dm Y  \) = - \f{\CN_S  }{8\pi} \U.
\eea
If we choose $\CN_S = \CA_S$, however, the classical non-chirality of this modified current exactly cancels the one-loop contribution\footnote{For some theories, further modification of currents is possible. Adding $i \f{\CN'_S}{4\pi} \Dm \bar Y$, for instance, works for theories in which $J^A$ and $E_A$ are independent of $Y$ as the EOMs give $\DBp \(\Dm \bar Y\) = 0$ . One example of this kind of theory is the HK \cite{Hori:2001} model.}.

\subsubsection{Stress Tensor}
Now that we have understood the $U(1)$ currents, let's move on to constructing the chiral stress tensor. 

The unimproved stress tensor can be obtained from N\"oether's procedure, however this is not gauge invariant.  Using integration by parts, the gauge invariant superfield completion of the stress tensor is:
\bea\label{barestress}
T_0^{++}  &=& - \f 12 \(\f{i}{4e^2} \U \p_- \bar \U +\sum_I e^{Q_I V} \Dm \Phi_I \Dm \bar \Phi_I + i \sum_A e^{q_A V } \bar{\Gamma}_{-A} \Dm \Gamma_{-A} + k^2 \Dm Y \Dm \bar Y \). \nonumber
\eea
This is equivalent to promoting the gravitational stress tensor to superspace.

Acting with $\DBp$ on $T_{0}^{++}$ and using the EOM and quasi-homogeneity conditions we get:
\bea
\DBp T_0^{++} = \f 12 \mu  \ \p_- \(\G_{-A} J^A \).
\eea
Thus $T_{0}^{++}$ is not quite chiral. We would like to improve it to a chiral stress tensor without ruining the conservation.

To achieve this, recall that a conserved current $\CJ ^\mu$ can be improved to $\CJ^{'\mu}$ by adding the divergence of an antisymmetric operator.
\bea
\CJ^{' \mu} = \CJ^\mu + \partial_{\nu}K^{[\mu\nu]} , \quad \rightarrow \quad \p_\mu \CJ^{' \mu} = \p_\mu \(\CJ^\mu +\partial_{\nu}K^{[\mu\nu]} \) = 0 
\eea
We want to find a $K^{[\mu\nu]}$ that renders $T^{++}$ chiral. In our case this will be of the form, $\epsilon^{\mu\nu}\partial_{\nu}\mathcal{F}$. 

As demonstrated by Silverstein and Witten \cite{Silverstein:1994}, given an R-symmetry, finding $K^{[\mu\nu]}$ is a simple matter. Consider the following, non-conserved, current:

\bea
\tilde J_R^+ = -i \(\sum_I e^{Q_I V} \alpha^R_I \Phi_I \Dm \bar \Phi_I +i \sum_A e^{q_A V} (\beta^R_A+1) \bar{\Gamma}_{-A} \Gamma_{-A} + k^2 \kappa^R \Dm \bar Y \).
\eea
Where $\alpha^R_I$ are the R-charges of the chiral scalars, $\beta^R_I$ are the R-charges of the left-moving fermions in the fermi multiplets, and $\kappa^R$ denotes the R-charge of the shift field. 

We emphasize that this is {\bf NOT} the R-current. This would be a regular chiral current for a global symmetry under which all components of each superfield transform with the same charge. However, as the quasi-homogeneity conditions corresponding to the R-symmetry are different from those for a standard global symmetry, the given current is not quite chiral.

\bea
\DBp  \tilde J_R^+  |_{EOMs} = - 2 \mu \ \Gamma_{-A} J^A,
\eea
In fact, $\tilde{J}^{+}_{R}$ is not chiral in exactly the correct way to compensate for the non-chirality of $T_{0}^{++}$. Now, we are able to construct a chiral stress tensor:
\bea
T^{++} \, &\equiv& \,  T_0^{++} + \frac{i}{4} \p_- \tilde J_R^+ \nonumber \\
&=&- \f 12 \(\f{i}{4e^2} \U \p_- \bar \U +\sum_I e^{Q_I V} \Dm \Phi_I \Dm \bar \Phi_I + i \sum_A e^{q_A V } \bar{\Gamma}_{-A} \Dm \Gamma_{-A} + k^2 \Dm Y \Dm \bar Y \) \nonumber \\
&& + \f 14 \p_-  \(\sum_I e^{Q_I V} \alpha^R_I \Phi_I \Dm \bar \Phi_I +i \sum_A e^{q_A V} (\beta^R_A+1) \bar{\Gamma}_{-A} \Gamma_{-A} + k^2 \kappa^R \Dm \bar Y \), \nonumber \\
\eea
such that
\bea
\DBp T^{++}   |_{EOMs} = 0 .
\eea
The lowest component is: 
\bea
t^{++} &=&- \f 12 \(\sum_I \Dm \phi_I  \Dm \bar \phi_I + i \sum_A \bar \gamma_{-A} \Dm \gamma_{-A} + k^2 \Dm y \Dm \bar y + \f {i}{2e^2} \lambda \Dm \bar \lambda \)  \nonumber \\
&& + \f 14 \p_- \(\sum_I \alpha^R_I \phi_I \Dm \bar \phi_I + i \sum_A (\beta^R_A+1) \bar \gamma_{-A} \gamma_{-A} + k^2 \kappa^R \Dm\bar y \),
\eea
which we identify as flowing to the IR stress tensor $\CT_L$.

To summarize, exploiting the existence of an R-symmetry and its quasi-homogeneity conditions, we are able to identify a chiral, gauge invariant, and left moving stress tensor.

\subsubsection{Operator Product Expansion for Chiral Operators}

Now that we have constructed candidate chiral operators, let's check the operator algebra. We consider a larger algebra, containing the gauge currents as well as the stress tensor and $U(1)_{L}$ current. The OPEs of the gauge current give the gauge and chiral anomalies. When calculating the Operator Product Expansion of the currents and stress tensor, we may use the free field OPEs for our component fields.

In the UV, where $ e \rightarrow 0 $, we can rescale our gauge field strength$ \U \rightarrow e \U$ to go to a free theory. The chiral operators in the free theory and the R-current are given below. We have not included the potential modifications of section (\ref{sec:mod currents}) because, as mentioned, they correspond to using the Green-Schwarz mechanism, which we address in the next section.

\bea
j_G^{+}  &=& - i \sum_I   Q_I   \phi_I \p_- \bar \phi_I  + \sum_A q_A \bar \gamma_{- A} \gamma_{- A } -  i k^2  \p_- \bar y  \nonumber  \\
j_L^+ &=& - i \sum_I \alpha^L_I \phi_I \p_- \bar \phi_I + \sum_A \beta^L_A \bar \gamma_{-A} \gamma_{-A} - i k^2 \kappa^L \p_-\bar y\nonumber\\
t^{++} &=&- \f 12 \(\sum_I \p_- \phi_I  \p_- \bar \phi_I + i \sum_A \bar \gamma_{-A} \p_- \gamma_{-A} + k^2 \p_- y \p_- \bar y + \f {i}{2} \lambda \p_- \bar \lambda \)  \nonumber \\
&& + \f 14 \p_- \(\sum_I \alpha^R_I \phi_I \p_- \bar \phi_I + i  \sum_A (\beta^R_A+1) \bar \gamma_{-A}  \gamma_{-A}  +  k^2 \kappa^R \p_-\bar y\)\nonumber\\
j_R^+ &=& -i \sum_I \alpha^R_I \phi_I \p_{-}\bar \phi_I +\sum_A \beta^R_A \bar \gamma_{-  A} \gamma_{- A}  -i  k^2 \kappa^R \p_{-}\bar y + \frac{1}{2} \bar \lambda_- \lambda_-  \nonumber\\
j_R^- &=& i\sum_I \(\alpha_I^R \phi_I \p_{+} \bar \phi_I - i \(\alpha_I^R-1\)\bar \psi_{+ I } \psi_{+ I }\) + i k^2 \kappa^R \p_{+}\bar y  - k^2 \bar \chi_+ \chi_+ 
\eea

The singular part of the operator product expansion for relevant operators are:
\bea
j_G^+(x) \, j_G^+(y) &\sim&  \frac{1}{(x^- -   y^-)^2} \(\sum_I Q_I  Q_I -  \sum_A q_A q_A\) \nonumber \\
j_G^+(x) \,   j_L^+(y) &\sim&  \frac{1}{(x^- -  y^-)^2} \(\sum_I Q _I  \alpha_I^L - \sum_A q_A \beta_A^L \)  \nonumber \\
j_L^+(x) \,  j_L^+(y) &\sim&   \frac{1}{(x^- -   y^-)^2} \(\sum_I \alpha_I^L  \alpha_I^L - \sum_A \beta_A^L \beta_A^L\) \nonumber  \\
j_R^+(x) \,  j_R^+(y) &\sim&   \frac{1}{(x^- -   y^-)^2} \(\sum_I \alpha_I^R  \alpha_I^R - \sum_A \beta_A^R \beta_A^R - \underbrace{1}_{\lambda_-} \) \nonumber\\
j_R^-(x) \,  j_R^-(y) &\sim&   \frac{1}{(x^+ -   y^+)^2} \(\sum_I \alpha_I^R  \alpha_I^R - \sum_I \(\alpha_I^R-1\)^2- \underbrace{1}_{\chi_+} \)  \nonumber \\
j_G^+(x) \,  t^{++} (y)&\sim&   \f{i}{2(x^- - y^-)^3}  \;  \(\sum_I Q_I \alpha_I^R - \sum_I Q_I - \sum_A q_A \beta_A^R \)-\frac{1}{2} \frac{ j_G^{+}}{\(x^- - y^-\)^2}-\frac{1}{2} \frac{\p_- j_G^{+}}{\(x^- - y^-\)}  \nonumber \\
j_L^+(x) \,  t^{++}(y) &\sim&   \frac{i}{2(x^- - y^-)^3} \;  \(\sum_I  \alpha_I^L  \alpha_I^R - \sum_I \alpha^L_I - \sum_A  \beta_A^L  \beta_A^R \)-\frac{1}{2}  \frac{j_L^{+}}{\(x^- - y^-\)^2} -\frac{1}{2}  \frac{\p_- j_L^{+}}{\(x^- - y^-\)}  \nonumber \\
t^{++}(x)  \,  t^{++}(y) &\sim&  \frac{1}{8 (x^- - y^-)^4} \; \(\sum_I \(3 (\alpha^R_I -1)^2 -1 \)  + \sum_A \(1- 3 (\beta^{R}_A)^2\) \)  \nonumber \\
&&-\frac{1}{2}\(\frac{t^{++}}{\(x^- - y^-\)^2}+\frac{\p_- t^{++}}{\(x^- - y^- \) }\)  
\eea

\subsubsection{Conditions for Conformality}

As mentioned previously, classically chiral operators can be anomalous.  In order to check the chirality at the quantum level one should investigate whether chirality holds within correlation functions. It can be shown that checking this is equivalent to the vanishing of the most singular terms in the gauge current OPE relations, see appendix \ref{GI Q Chirality}. Requiring the existence of a chiral, $(0,2)$ superconformal algebra yields the following anomaly cancellation conditions: 
\bea
U(1)_G  \, U(1)_G :&&  \sum_I  Q_I   Q_I  - \sum_A q_A  q_A = 0  \nonumber \\
U(1)_G \, U(1)_R:&&  \sum_I Q _I    \(\alpha_I^R - 1\)- \sum_A q_A  \beta_A^R = 0 \nonumber \\
U(1)_G \, U(1)_L:&& \sum_I  Q_I \alpha_I^L - \sum_A q_A   \beta_A^L =0 \nonumber \\
U(1)_L \, U(1)_R:&& \sum_I \alpha_I^L  \(\alpha_I^R - 1\)-\sum_A \beta_A^L  \beta_A^R = 0. \
\eea
The charge and rank of the vector bundle in the IR theory may also be gleaned from the leading singularities of the OPEs.
\bea
\CJ_L  (x)\; \CJ_L (y) &\sim& \frac{r_L}{\(x^- - y^-\)^2}\\
\CT_L (x)  \;  \CT_L(y) &\sim& \frac{1}{2} \frac{c_L}{\(x^- - y^-\)^4} + \frac{2}{\(x^- - y^-\)^2} \CT(y) + \frac{1}{\(x^- - y^-\)} \p_- \CT(y) \\
\CJ_R  (x)\; \CJ_R (y) &\sim& \frac{1}{2}\frac{\hat c_R}{ \(x^+ - y^+\)^2}.
\eea
The last equation is equivalent to calculating the leading coefficient of the $j_R^+ j_R^+$ OPE and subtracting the leading coefficient of the $j_R^- j_R^-$ OPE. This gives one third of the right moving central charge $\hat c_R = \frac 13 c_R$.  It is conventional to normalize $\CT_L = -2  \, t^{++}$.  Then the UV OPEs tell us that: 
\bea
c_L &=& \sum_I \(3 (\alpha^R_I -1)^2 -1 \)  + \sum_A \(1- 3 (\beta^{R}_A)^2\)\\
r_L &=&  \sum_I \alpha_I^L  \alpha_I^L - \sum_A  \beta_A^L  \beta_A^L\\
\hat c_R &=& \sum_I \(\alpha_I^R -1 \)^2 -\sum_A \beta^R_A \beta^R_A.
\eea

Note, $c_{L}-c_{R}=\sum_{A}1-\sum_{I}1$, which is an RG invariant quantity as expected.

\subsection{Anomalous Model with Green-Schwarz Mechanism}

Now that we have warmed up with the classically gauge invariant case, lets consider adding to the action the following non-gauge invariant piece, to incorporate the Green-Schwarz mechanism.
\bea\label{EQ:ActionGS}
 \CL_{GS} = - \f{\hat \CC}{4} \int  d \tp Y \U + h.c. 
\eea
Here we take $\hat \CC$ to be real which represents a convention for the periodicity of the imaginary part of $Y$. The motivation behind adding this term is that this classical, gauge variant piece will cancel against the quantum anomaly. This is reasonable as the classical variation of the added term has exactly the same form as an anomaly. 

We follow the same procedure as before, but now with the GS term. The equations of motion are slightly modified.  

\bea
\DBp \(e^{q_A V} \bar{\Gamma}_{-A} \) &=&- \mu \  J^A \\
\DBp \(e^{Q_I V} \Dm \bar \Phi_I \) &=&- i \mu \sum_A \Gamma_{-A} \p_I J^A + i \sqrt{2} \sum_A e^{q_A V} \bar{\Gamma}_{-A} \p_I E_A  \\
k^{2}\DBp \(\Dm \bar Y \) &=&  i \mu \sum_A \Gamma_{-A} J^{'A} e^{-Y} - i \sqrt{2} \sum_A e^{q_A V}\bar{\Gamma}_{-A} E'_A e^{-Y}  - \f{i \hat \CC }{2} \U \\
  \f{1}{2e^2} \DBp\p_- \bar \U &=& \sum_I e^{Q_I V}Q_I \Phi_I \Dm \bar \Phi_i + i \sum_A e^{q_A V} q_A \bar{\Gamma}_{-A} \Gamma_{-A} + k^2 \Dm \bar Y  -\hat \CC  \, \p_- Y 
\eea 

It is interesting to note that acting with $\DBp$ on the last equation and using the quasi-homogeneity conditions reveals: 
\bea
\hat \CC \, \U = 0. 
\eea 
When the shift field is charged and the GS mechanism is in play, the field strength multiplet vanishes on-shell. 

\subsubsection{Modification of Currents: Stress tensor and $U(1)$ currents}
The action is no longer gauge invariant. As a consequence, the canonical currents transform under the gauge variation. All is not lost, however. Thanks to the shift field we may improve the currents to be gauge invariant.

Let us consider the contribution from the GS term to the currents. An interesting feature of the GS term is that, though it destroys classical gauge invariance, it preserves the global part of the symmetry. The action shifts by a total derivative.

Recall that when the action changes by a total derivative under a symmetry transformation, 
\bea
\delta \CL(x) = \p_\mu K^\mu, \nonumber
\eea 
the conservation of currents get shifted.
\bea
\p_\mu j^\mu(x) = \delta \CL(x) - \f{\delta S}{\delta \phi_a(x)} \delta \phi_a (x)  = \p_\mu K^\mu(x)  - \f{\delta S}{\delta \phi_a(x)} \delta \phi_a (x) \nonumber
\eea
\bea
\p_\mu \(j^\mu (x) - K^\mu(x)\) = 0.
\eea

The shift field transforms as $y \rightarrow y - i \kappa$ and the GS term leads to the variation:
\bea
\delta \CL_{GS}  = -\frac{1}{2} \hat \CC  \kappa (\p_+ v_-  - \p_- v_+) , 
\eea
implying $K^+ = - \frac 12 \hat \CC \kappa v_-$.  Thus the effect of the GS term on the currents is:
\bea
j^+ \rightarrow j^+ + \f 12 \hat \CC  \kappa v_-,
\eea
which is not gauge invariant. This is plausible as the GS term is not gauge invariant. Luckily, as mentioned, the shift field allows us to improve the current to a gauge invariant form.
\bea
j^+ &\rightarrow& j^+ -   i \hat \CC\kappa \p_-  y 
\eea
The supersymmetric completion of this is:
\bea
J^+ \rightarrow J^+ - i \hat \CC \kappa \Dm Y. 
\eea
Notice that this is exactly the form of the current modification in \ref{sec:mod currents}.
Thus with the GS contribution our supersymmetric currents can be written as,
\bea
J_G^+ = -i \(\sum_I e^{Q_I V} Q_I \Phi_I \Dm \bar \Phi_I +i \sum_A e^{q_A V} q_A\bar{\Gamma}_{-A} \Gamma_{-A} + k^2 \Dm \bar Y + \hat \CC  \, \Dm Y \)
\eea
\bea
J_L^+ = -i \(\sum_I e^{Q_I V} \alpha^L_I \Phi_I \Dm \bar \Phi_I +i \sum_A e^{q_A V} \beta^L_A \bar{\Gamma}_{-A} \Gamma_{-A} + k^2 \kappa^L \Dm \bar Y + \kappa \hat \CC \Dm Y \). 
\eea
These are gauge invariant and chiral, as before.

A little more care is required to construct the stress tensor. Recall that the canonical stress tensor is not always gauge invariant. In fact, it can in general only be improved to be gauge invariant on shell. Varying with respect to the metric, however, produces a gauge invariant, symmetric, conserved stress tensor. Indeed this is one way to derive the expression for $T_{0}^{++}$, (\ref{barestress}). This method of constructing the stress tensor is particularly nice. The Green Schwarz term is metric-independent and so does not contribute to this definition of the stress tensor. The improvement term does change. Just as for the regular currents we have:
\begin{eqnarray}
\tilde{J}^{+}_{R}&\rightarrow&\tilde{J}^{+}_{R}-i\frac{\kappa^{R}\hat{\mathcal{C}}}{\mathcal{Q}}\mathcal{D}_{-}Y.
\end{eqnarray}
Putting this together yields an expression for $T^{++}$.
\begin{eqnarray}
T^{++}&=&- \f 12 \(\f{i}{4e^2} \U \p_- \bar \U +\sum_I e^{Q_I V} \Dm \Phi_I \Dm \bar \Phi_I + i \sum_A e^{q_A V } \bar{\Gamma}_{-A} \Dm \Gamma_{-A} + k^2 \Dm Y \Dm \bar Y \) \nonumber \\
&& + \f 14 \p_-  \(\sum_I e^{Q_I V} \alpha^R_I \Phi_I \Dm \bar \Phi_I +i \sum_A e^{q_A V} (\beta^R_A+1) \bar{\Gamma}_{-A} \Gamma_{-A} + k^2 \kappa^R \Dm \bar Y +\frac{\kappa^{R}\hat{\mathcal{C}}}{\mathcal{Q}}\mathcal{D}_{-}Y\) \nonumber \\
\end{eqnarray}
The chirality of this expression is ensured by the vanishing of the field strength multiplet.

\subsubsection{Conditions for Conformality}
With the improved gauge invariant currents, we can obtain constraints on the existence of an IR conformal algebra. As in the gauge invariant case (\ref{GI Q Chirality}), we check that the chirality condition is not anomalous. The details of the calculation are relegated to Appendix (\ref{opes}), but the results are presented below.  For  our model with GS term to flow into a $(0,2)$ superconformal field theory, we require the following anomaly cancellation conditions:
\bea\label{EQ:ConfGS}
U(1)_G  \, U(1)_G :&& \sum_I  Q_I   Q_I  - \sum_A q_A  q_A -2\hat\CC  = 0  \nonumber \\
U(1)_G \, U(1)_R:&& \sum_I  Q _I  \(\alpha_I^R - 1\) - \sum_A q_A  \beta_A^R  - 2 \hat \CC \kappa^R= 0  \nonumber \\
U(1)_G \, U(1)_L:&& \sum_I  Q_I \alpha_I^L - \sum_A q_A   \beta_A^L - 2 \hat\CC \kappa^L =0 \nonumber \\
U(1)_L \, U(1)_R:&& \sum_I \alpha_I^L  \(\alpha_I^R - 1\)-\sum_A \beta_A^L  \beta_A^R - 2 \hat\CC \kappa^L \kappa^R= 0. \
\eea
When there is a non trivial IR theory, the central charges and the rank of the vector bundle can also be identified, including the modifications from the GS term, this gives: 
\bea
c_L &=& \sum_I \(3 (\alpha^R_I -1)^2 -1 \)  + \sum_A \(1- 3 (\beta^{R}_A)^2\) - 6\hat \CC (\kappa^R)^{2}  \\
r_L &=&  \sum_I \alpha_I^L  \alpha_I^L - \sum_A  \beta_A^L  \beta_A^L - 2\hat \CC (\kappa^L)^{2} \\
\hat c_R &=& \sum_I \(\alpha_I^R -1 \)^2 -\sum_A \beta^R_A \beta^R_A - 2 \hat \CC (\kappa^R)^2.
\eea

\subsection{Multiple $U(1)$s}

Now that we have warmed up with a single gauge group and shift field, we are ready to take on the more general case. The action is written in equations (\ref{FullLag}-\ref{SFLagend}), but we repeat it below.
\begin{eqnarray}
\mathcal{L}&=&\mathcal{L}_{gk}+\mathcal{L}_{1}+\mathcal{L}_{W}+\mathcal{L}_{GS}+\mathcal{L}_{FI}
\end{eqnarray}
Where:
\begin{eqnarray}
\mathcal{L}_{gk}&=&-\frac{1}{2}\sum_{a}\int d^{2}\theta^{+}\frac{1}{4e^{2}_{a}}\bar{\Upsilon}_{a}\Upsilon_{a}\\
\mathcal{L}_{1}&=&-\frac{1}{2}\int d^{2}\theta^{+}\left(i\sum_{I}e^{Q_{I}^{a}V_{a}}\bar{\Phi}_{I}\mathcal{D}_{-}\Phi_{I}+i\sum_{l}\frac{k^{2}_{l}}{2}(Y_{l}+\bar{Y}_{l}+\mathcal{Q}_{l}^{a}V_{a})\mathcal{D}_{-}(Y_{l}-\bar{Y}_{l})+\sum_{A}e^{q_{A}^{a}V_{a}}\bar{\Gamma}_{-A}\Gamma_{-A}\right)\nonumber\\
\\
\mathcal{L}_{W}&=&-\frac{\mu}{2}\int d\theta^{+}\sum_{A}\Gamma_{-A}J^{A}+h.c.\\
\mathcal{L}_{GS}&=&\frac{1}{4}\int d^{2}\theta^{+}\sum_{a,b,l}\hat{\mathcal{C}}^{[a}_{l}\mathcal{Q}^{b]l}V_{a}A_{b}-\frac{1}{4}\sum_{a,l}\hat{\CC}^{al} \int d\theta^{+}\Upsilon_{a} Y_{l}+h.c.\\
\mathcal{L}_{FI}&=&-\frac{1}{4}\sum_{a}t^{a} \int d\theta^{+}\Upsilon_{a}+h.c.
\end{eqnarray}

\subsubsection{Equations of Motion}
The equations of motion differ only by the addition of extra indices.
\begin{eqnarray}
\bar{D}_{+}\left(e^{q_{A}^{a}V_{a}}\bar{\Gamma}_{-A}\right)&=&-\mu \ J^{A}\\
\bar{D}_{+}\left(e^{Q_{I}^{a}V_{a}}\mathcal{D}_{-}\bar{\Phi}_{I}\right)&=&-i\mu\sum_{A}\Gamma_{-A}\partial_{I}J^{A}+i\sqrt{2}\sum_{A}e^{q_{A}^{a}V_{a}}\bar{\Gamma}_{-A}\partial_{I}E_{A}\\
k_{l}^{2}\bar{D}_{+}\left(\mathcal{D}_{-}\bar{Y}_l\right)&=&-i\mu\sum_{A}\Gamma_{-A}\partial_{l}J^{A}+i\sqrt{2}\sum_{A}e^{q_{A}^{a}V_{a}}\bar{\Gamma}_{-A}\partial_{l}E_{A}-\frac{i}{2}\sum_{a}\hat{\mathcal{C}}^{al}\Upsilon_{a}\\
\frac{1}{2e_{a}^{2}}\left(\bar{D}_{+}\bar{\Upsilon}_{a}-D_{+}\Upsilon_{a}\right)&=&\sum_{I}e^{Q_{I}^{b}V_{b}}Q_{I}^{a}|\Phi_{I}|^{2}+\sum_{l}(k_{l}^{2}\mathcal{Q}_{l}^{a}-\hat{\mathcal{C}}^{al})(Y_{l}+\bar{Y}_{l}+\mathcal{Q}_{l}^{b}V_{b})+\sum_{b,l}\hat{\mathcal{C}}^{(a}_{l}\mathcal{Q}^{b)l}V_{b}\nonumber\\
\\
-\frac{1}{2e_{a}^{2}}\left(\bar{D}_{+}\partial_{-}\bar{\Upsilon}_{a}+D_{+}\partial_{-}\Upsilon_{a}\right)&=&\sum_{I}e^{Q_{I}^{b}V_{b}}Q_{I}^{a}\bar{\Phi}_{I}\overleftrightarrow{\mathcal{D}}_{-}\Phi_{I}+\sum_{l}(k_{l}^{2}\mathcal{Q}_{l}^{a}+\hat{\mathcal{C}}^{al})\mathcal{D}_{-}(Y_{l}-\bar{Y}_{l})-i\sum_{b,l}\hat{\mathcal{C}}^{(a}_{l}\mathcal{Q}^{b)l}A_{b}\nonumber\\
&-&2i\sum_{A}e^{q_{A}^{b}V_{b}}q_{A}^{a}\bar{\Gamma}_{-A}\Gamma_{-A}
\end{eqnarray}
Combining the last two of these yields:
\begin{eqnarray}
\frac{1}{2e_{a}^{2}}\bar{D}_{+}\partial_{-}\bar{\Upsilon}_{a}&=&\sum_{I}e^{Q_{I}^{b}V_{b}}Q_{I}^{a}\Phi_{I}\mathcal{D}_{-}\bar{\Phi}_{I}+\sum_{l}\(k_{l}^{2}\mathcal{Q}_{l}^{a}\mathcal{D}_{-}\bar{Y}_{l}- \hat \CC^{al} \Dm Y_l\)+\frac{i}{2}\sum_{b,l}\hat{\mathcal{C}}^{(a}_{l}\mathcal{Q}^{b)l}(A_{b}-i\partial_{-}V_{b})\nonumber \\
&&+i\sum_{A}e^{q_{A}^{b}V_{b}}q_{A}^{a}\bar{\Gamma}_{-A}\Gamma_{-A}
\end{eqnarray}

Notice that acting with $\DBp$  on the above equation, we find:
\bea
0 &=&  \sum_{b, l} \hat \CC^{(a}_l \CQ_l^{b)} \U_b 
\eea

\subsubsection{Modification of $U(1)$ currents}
Now let us look at the various currents. Constructing the gauge invariant conserved currents is a little more involved. The results are:
\bea
 i J^{a +}_G = \sum_I e^{Q^b_I V_b} Q^a_I \Phi_I \Dm \bar \Phi_I +i \sum_A e^{q^b_A V_b} q^a_A\bar{\Gamma}_{-A} \Gamma_{-A} + \sum_l \(k_l^2 \CQ_l^a \Dm \bar Y_l + \mathcal{U}^{al}\partial_{-}Y_{l}+\frac{i}{2}\hat{\mathcal{C}}^{bl}\mathcal{Q}^{a}_{l}(A_{b}-i\partial_{-}V_{b})\) \nonumber
\eea
\bea
i J_L^+ = \sum_I e^{Q^a_I V_a} \alpha^L_I \Phi_I \Dm \bar \Phi_I +i \sum_A e^{q^a_A V_a} \beta^L_A \bar{\Gamma}_{-A} \Gamma_{-A} +\sum_l \( k_{l}^2 \kappa_l^L \Dm \bar Y_l + \mathcal{U}^{Ll}\partial_{-}Y_{l}+\frac{i}{2}\hat{\mathcal{C}}^{bl}\kappa^{L}_{l}(A_{b}-i\partial_{-}V_{b}) \). \nonumber
\eea
Here the $\mathcal{U}$s are chosen to make the various currents gauge invariant:
\begin{eqnarray}
\sum_l \mathcal{U}^{al}\mathcal{Q}^{b}_{l}=\sum_l \hat{\mathcal{C}}^{bl}\mathcal{Q}^{a}_{l}\\
\sum_l \mathcal{U}^{Ll}\mathcal{Q}^{b}_{l}=\sum_l \hat{\mathcal{C}}^{bl}\kappa^{L}_{l}.
\end{eqnarray}

It is not true that every model admits $\mathcal{U}$s satisfying these conditions. Consider, for example, a model consisting of two $U(1)$s, with a single shift charged field $Y$, charged under a single factor, $\mathcal{Q}^{a}=(0,2)$, with a particular GS term specified by $\hat{C}^{a}=(1,0)$. This gives the symmetric anomaly matrix and anti-symmetric matrix:
\begin{eqnarray}
\hat{\mathcal{C}}^{(a}\mathcal{Q}^{b)}&=&\left(\begin{array}{c c}
0 & 1 \\
1 & 0
\end{array}\right)\nonumber\\
\hat{\mathcal{C}}^{[a}\mathcal{Q}^{b]}&=&\left(\begin{array}{c c}
0 & 1 \\
-1 & 0
\end{array}\right).
\end{eqnarray}
Bosonic component fields of the GS Lagrangian can be written as:
\begin{eqnarray}
\mathcal{L}_{GS}=\theta\epsilon^{\mu\nu}F_{1\mu\nu}-2\epsilon^{\mu\nu}A_{1\mu}A_{2\nu}.
\end{eqnarray}
The variation of this term is symmetric and can be cancelled off of otherwise anomalous fermion content, rendering the total anomaly zero. Despite this apparent good behavior, the term contributes to the U(1) currents in a gauge variant fashion.
\begin{eqnarray}
j_{1}^{\mu}&\propto&\theta\epsilon^{\mu\nu}A_{1\nu}+\ldots
\end{eqnarray}
Since $Y$ doesn't shift under the first $U(1)$, this cannot be improved to a gauge invariant expression. There exists no $\mathcal{U}$ for this theory.

The existence of gauge invariant currents imposes non-trivial constraints on the matrix $\mathcal{C}^{al}$. Of course, if $\mathcal{Q}$ is invertible, i.e. if there is a shift charged field for every $U(1)$ factor, then $\mathcal{U}^{a}$ and $\mathcal{U}^{L}$ exist. These constrains should imply non-trivial restrictions on the types of geometries that can be realized via $(0,2)$ Green-Schwarz GLSMs. It would be interesting to understand the larger implications of these equations and whether they have a natural mathematical interpretation.

The stress tensor is defined in a similar fashion as before, however now the existence of a gauge invariant R current requires a $\mathcal{U}^{R}$. The stress tensor is given by:
\begin{eqnarray}
&&T^{++}=- \f 12 \(\sum_{a}\f{i}{4e^2} \U_{a} \p_- \bar \U_{a} +\sum_I e^{Q_{I}^{a} V_{a}} \Dm \Phi_I \Dm \bar \Phi_I + i \sum_A e^{q_{A}^{a} V_{a} } \bar{\Gamma}_{-A} \Dm \Gamma_{-A} + \sum_{l}k_{l}^2 \Dm Y \Dm \bar Y \) \nonumber \\
&&+ \f 14 \p_-  \(\sum_I e^{Q_{I}^{a}V_{a}} \alpha^R_I \Phi_I \Dm \bar \Phi_I +i \sum_A e^{q_{A}^{a}V_{a}} (\beta^R_A+1) \bar{\Gamma}_{-A} \Gamma_{-A} + \sum_{l}k_{l}^{2} \kappa_{l}^{R} \Dm \bar Y\right.\nonumber\\
&&+\left. \mathcal{U}^{Rl}\partial_{-}Y_{l}+\frac{i}{2}\hat{\mathcal{C}}^{bl}\kappa^{R}_{l}(A_{b}-i\partial_{-}V_{b})\).
\end{eqnarray}
With:
\begin{eqnarray}
\sum_l \mathcal{U}^{Rl}\mathcal{Q}^{b}_{l}=\sum_l \hat{\mathcal{C}}^{bl}\kappa^{R}_{l}.
\end{eqnarray}
$T^{++}$ is manifestly gauge invariant. Chirality follows from the equations of motion and the quasi-homogeneity conditions.
\subsubsection{Conditions on Quantum Chirality: Multiple $U(1)$s and Shift Fields}
When there exist improved gauge invariant currents, we can use them to obtain constraints on the existence of an IR conformal algebra.  The details of the calculation are in Appendix (\ref{opes}) and only the results are presented below.

The anomaly cancellation conditions for multiple $U(1)$s and shift fields are:
\bea
U(1)^a_G  \, U(1)^b_G :&& \sum_I  Q^a_I   Q^b_I  - \sum_A q^a_A  q^b_A - 2 \sum_l \CU_l^{(a}\CQ_l ^{b)}   = 0  \nonumber \\
U(1)^a_G \, U(1)_R:&&   \sum_I Q^a_I  \alpha_I^R - \sum_I Q^a_I - \sum_A q^a_A \beta_I^R - 2 \sum_l \hat \CC^a_l \kappa_l^R= 0  \nonumber \\
U(1)^a_G \, U(1)_L:&& \sum_I  Q^a_I \alpha_I^L - \sum_A q^a_A   \beta_A^L - \sum_l \(\CU^{al} \kappa^L_l + \CU^{L l } \CQ^a_l \) =0 \nonumber \\
U(1)_L \, U(1)_R:&& \sum_I \alpha_I^L \alpha_I^R - \sum_I \alpha^L_I - \sum_A \beta_A^L  \beta_A ^R   - \sum_l\(  \CU^{L l} \kappa^R_l + \CU^{R l} \kappa^L_l\) = 0.\
\eea
One can also calculate parameters of the CFT and central charges get contribution from multiple gauge generators and shift fields
\bea
c_L &=&\sum_I \(3 (\alpha^R_I -1)^2 -1 \)  + \sum_A \(1- 3 (\beta^{R }_A)^2\) - 6 \sum_l \CU^{Rl} \kappa^R_l  + \sum_l 2 - \sum_a 2  \\
r_L &=& \sum_I \alpha_I^L  \alpha_I^L - \sum_A  \beta_A^L  \beta_A^L - 2\sum_l  \CU^{L l } \kappa^L_l  \\
\hat c_R &=&\sum_I \(\alpha_I^R - 1\)^2- \sum_A  \beta_A^R  \beta_A^R - \sum_a 1 - 2\sum_l  \CU^{R l } \kappa^R_l+\sum_l 1
 \eea

Again $c_{L}-c_{R}$ is manifestly invariant.

%
%
%
\section{Conclusions}\label{Sec:Conclusions}

In this paper, we have argued that \nK\ geometries with $H$-flux satisfying the modified Bianchi identity can be described by standard \ZT\ GLSMs without any additional structure.  
At low energies, the non-trivial $H$-flux is realized through a set of effective Green-Schwarz terms canceling an anomaly in the fermion measure.  These Green-Schwarz terms arise by integrating out charged fermions which are massive in the local patch of the moduli space; the Green-Schwarz terms ensure the cancellation of the total anomaly after truncating to the (would-be anomalous) spectrum of surviving light fermions.
This clarifies, for example, how $dH\neq0$ arises in \ZT\ deformations of \TT\ models corresponding to deformations of the vector bundle away from standard embedding.  
This improved understanding of the familiar \ZT\ GLSM then allows us to realize various previously-constructed quasi-linear models for non-K\"ahler manifolds with torsion \cite{Adams:2006,AdamsGuarrera:2009,Quigley:2011,Blaszczyk:2011}\ as effective descriptions of certain patches of the moduli space of elementary \ZT\ GLSMs.  

As an independent check of the consistency of these effective descriptions, we studied the quantum consistency of a simple class of such quasi-linear models involving an anomalous gauge group together with anomaly canceling Green-Schwarz axions.  Effectively, this linearizes the GS sector.  This allowed us to show, following \cite{Silverstein:1994,AdamsLapan:2009},  that GS anomaly cancellation in these models ensures the existence of an off-shell $\CN=2$ superconformal multiplet whose OPE algebra closes correctly within $\QPB$-cohomology.  We then used these OPEs to compute the central charge of the \ZT\ SCFT to which the theory is expected to flow in the IR.  

Many questions remain about the geometry and moduli space of such \ZT\ GLSMs and their quasi-linear effective descriptions.  At a technical level, it would be reassuring to explicitly derive the equivalence between the NLSM and GLSM completions of models with ${\rm tr}R\wedge R\neq {\rm Tr}F\wedge F$  by computing the full one-loop effective action of the GLSM and verifying that it is in the same universality class as the standard NLSM construction of \cite{Hull:1985,Sen:1985}.  It would be surprising if these construction do not agree, since that would give us a new way to complete naively anomalous string theories to good compactifications.

Much more interesting are questions about the global moduli space of generic $(0,2)$ GLSMs.   What, in the GLSM, distinguishes models which topologically admit a \Ka\ structure (such as small deformations of \TT\ models) from models which do not (such as the \nK\ $T^{2}$-fibration GLSMs of \cite{Adams:2006})? Relatedly, when can two such models (one admitting a \Ka\ structure, one not) be realized as different phases of a single underlying GLSM connected by a smooth quantum transition?  More generally, what is the global moduli space of a \ZT\ GLSM and does it contain multiple inequivalent \TT\ sub-loci which are embedded smoothly? To this end, it would also be of great interest to be able to compute topological invariants for specific non-K\"ahler manifolds using the structure of the $(0,2)$ GLSM.

Another interesting question is whether worldsheet duality along the lines of \cite{Rocek:1991,Morrison:1995,Hori:2000,Adams:2003}\ might be applicable to the GS axion models.  On the surface it would seem less than useful -- under such an abelian duality, axial couplings of a scalar to a gauge field are exchanged with canonical couplings, but in the present models our GS scalars are both axially and canonically coupled.  However, the role of the anomaly may alter the naive dualization.  Even if the anomaly just goes along for the ride, however, such dualities may prove helpful in patching together a clear picture of the moduli space of these theories.

There are clearly many interesting geometric questions waiting to be attacked through more careful analysis of the torsional moduli space of \ZT\ GLSMs. We hope to return to them soon. \\

{\bf Note Added:} During the completion of this text, a paper with some overlap with our observations appeared on the arxiv, \cite{Quigley:2012}.

%
%
\section*{Acknowledgments}

We thank 
Jacques Distler, 
Kentaro Hori, 
Joshua Lapan, 
Ilarion Melnikov, 
John McGreevy, 
David Morrison, 
Stefan Nibbelink, 
Daniel Park
and
Callum Quigley, 
Sav Sethi 
for discussions.
A.A. also thanks the organizers and participants of the 2011 ESI Workshop on $(0,2)$ Mirror Symmetry for providing a stimulating environment where some of these ideas were discussed, and the 2012 Simons Workshop on String Theory for Mathematicians and the Isaac Newton Institute for hospitality during the final editing of this note.
The work of A.A is supported in part by funds provided by the U.S. Department of Energy (D.O.E.) under cooperative research agreement \#DE-FC02-94ER40818.
The work of J.L. is supported in part by the Samsung Scholarship. 

%
%
\appendix
%

%
%
\section{Conventions}\label{Sec:Conventions}

\subsection{Coordinates}
We work in $1+1$ space-time dimensions with coordinates $x^{0}$, $x^{1}$ and Lorentzian metric. For most of the calculations we use light-cone coordinates. The relevant formulae are:
\begin{eqnarray}
x^{\pm}&=&\frac{1}{2}(x^{0}\pm x^{1}), \qquad \partial_{\pm}=\partial_{0}\pm\partial_{1}, \qquad g_{(lc)}^{+-}=\epsilon^{+ -}= -\frac{1}{2}.
\end{eqnarray}
We have superspace coordinates, $\theta^{+}$, and $\bar{\theta}^{+}$. Integrals are normalized as:
\begin{eqnarray}
\int d^{2}\theta\theta^{+}\bar{\theta}^{+}&=&\int d\theta^{+}\theta^{+}=-\int d\bar{\theta}^{+}\bar{\theta}^{+}=1.
\end{eqnarray}
We also define superspace operators:
\begin{eqnarray}
Q_{+}&=&\frac{\partial}{\partial\theta^{+}}+i\bar{\theta}^{+}\partial_{+},  \qquad \bar{Q}_{+}=-\frac{\partial}{\partial\bar{\theta}^{+}}-i\theta^{+}\partial_{+} \\
D_{+}&=&\frac{\partial}{\partial\theta^{+}}-i\bar{\theta}^{+}\partial_{+},   \qquad \bar{D}_{+}=-\frac{\partial}{\partial\bar{\theta}^{+}}+i\theta^{+}\partial_{+}.\nonumber
\end{eqnarray}
These satisfy:
\begin{eqnarray}
D_{+}^{2}&=&\bar{D}_{+}^{2}=0\\
Q_{+}^{2}&=&\bar{Q}_{+}^{2}=0\nonumber\\
\{D_{+},\bar{D}_{+}\}&=&2 i \partial_{+}\nonumber\\
\{Q_{+},\bar{Q}_{+}\}&=&-2 i \partial_{+}\nonumber\\
\{Q_{+},D_{+}\}&=&\{Q_{+},\bar{D}_{+}\}=0 .\nonumber
\end{eqnarray}
When symmetrizing and anti-symmetrizing indices we take:
\bea
M_{[a,b]} = \frac 12 \(M_{ab} - M_{ba}\), \quad M_{(a,b)} = \frac 12\(M_{ab}+ M_{ba}\) 
\eea

\subsection{Superfields}
A chiral superfield, $\Phi$, satisfies $\bar{D}_{+}\Phi=0$. A Fermi superfield, $\Gamma$, satisfies $\bar{D}_{+}\Gamma=\sqrt{2}E$. These have component Expansion:
\begin{eqnarray}
\Phi&=&\phi+\sqrt{2}\theta^{+}\psi_{+}-i\theta^{+}\bar{\theta}^{+}\partial_{+}\phi\\
\Gamma&=&\gamma-\sqrt{2}\theta^{+}G-\sqrt{2}\bar{\theta}^{+}E-i\theta^{+}\bar{\theta}^{+}\partial_{+}\gamma.
\end{eqnarray}
We also use a shift charged chiral field, $Y$. We take it to have the component expansion:
\begin{eqnarray}
Y&=&y+\sqrt{2}\theta^{+}\chi_{+}-i\theta^{+}\bar{\theta}^{+}\partial_{+}y.
\end{eqnarray}
In addition to the chiral fields there are gauge fields. As apposed to the $(2,2)$ case, which requires a single superfield, there are two $(0,2)$ vector fields, $A$, $V$. In Wess-Zumino gauge these have the component expansion.
\begin{eqnarray}
V&=&\theta^{+}\bar{\theta}^{+}v_{+}\\
A&=&v_{-}+\sqrt{2}i\theta^{+}\bar{\lambda}_{-}+\sqrt{2}i\bar{\theta}^{+}\lambda_{-}+2\theta^{+}\bar{\theta}^{+}D.
\end{eqnarray}
From this we can construct the fermionic, gauge invariant field strength, $\Upsilon$.
\begin{eqnarray}
\Upsilon&=&\bar{D}_{+}(A-i\partial_{-}V)\\
&=&-i\sqrt{2}\left(\lambda_{-}+\sqrt{2}\theta^{+}(F_{01}+iD)-i\theta^{+}\bar{\theta}^{+}\partial_{+}\lambda_{-}\right)\nonumber
\end{eqnarray}
$F_{01}= \p_0 v_1 - \p_1 v_0 $ is the gauge field strength.

We have both conventionally charged fields, $\Phi$, $\Gamma$, and shift charged fields, $Y$. The gauge transformations with chiral gauge parameter $\Lambda$ are summarized bellow.
\begin{eqnarray}
\Phi&\rightarrow&e^{-iQ\Lambda}\Phi\\
\Gamma&\rightarrow&e^{-iq\Lambda}\Gamma\nonumber\\
Y&\rightarrow&Y-i\mathcal{Q}\Lambda\nonumber\\
V&\rightarrow&V+i(\Lambda-\bar{\Lambda})\nonumber\\
A&\rightarrow&A+\partial_{-}(\Lambda+\bar{\Lambda})\nonumber\\
\end{eqnarray}
The different qs, $Q$, $q$, and $\mathcal{Q}$ represent our naming conventions for the charges of chiral, fermi, and shift fields respectively. To facilitate writing down a gauge invariant action in superspace, it is convenient to define the following covariant derivatives:
\begin{eqnarray}
\mathcal{D}_{-}\Phi&=&\left(\partial_{-}+\frac{Q}{2}(\partial_{-}V+iA)\right)\Phi\\
\mathcal{D}_{-}\Gamma&=&\left(\partial_{-}+\frac{q}{2}(\partial_{-}V+iA)\right)\Gamma\\
\mathcal{D}_{-}Y&=&\partial_{-}Y+\frac{\mathcal{Q}}{2}(\partial_{-}V+iA).\nonumber
\end{eqnarray}
Lastly, for the FI parameter we use
\bea
t = r + i \theta, \qquad \theta \sim \theta+ 2\pi.
\eea


\section {Action}\label{actions}
Here, for completeness, we record the full action we use for the shift charged field, and the complete component expansion for the gauge invariant case.

The action is given by:
\begin{eqnarray}\label{FullLag}
\mathcal{L}&=&\mathcal{L}_{gk}+\mathcal{L}_{m}+\mathcal{L}_{W}+\mathcal{L}_{GS}+\mathcal{L}_{FI}
\end{eqnarray}
where $\mathcal{L}_{gk}$ contains the gauge kinnetic terms, $\mathcal{L}_{m}$ contains the kinetic terms for the matter as well as the gauge interactions. $\mathcal{L}_{W}$ is the superpotential, $\mathcal{L}_{GS}$ is the Green-Schwarz term , and the $\mathcal{L}_{FI}$ is the conventional Fayet-Iliopoulos(FI) term.
\begin{eqnarray}
\mathcal{L}_{gk}&=&-\frac{1}{2}\sum_{a}\int d^{2}\theta^{+}\frac{1}{4e^{2}_{a}}\bar{\Upsilon}_{a}\Upsilon_{a}\\
\mathcal{L}_{m}&=&-\frac{1}{2}\int d^{2}\theta^{+}\left(i\sum_{I}e^{Q_{I}^{a}V_{a}}\bar{\Phi}_{I}\mathcal{D}_{-}\Phi_{I}+i\sum_{l}\frac{k^{2}_{l}}{2}(Y_{l}+\bar{Y}_{l}+\mathcal{Q}_{l}^{a}V_{a})\mathcal{D}_{-}(Y_{l}-\bar{Y}_{l})+\sum_{A}e^{q_{A}^{a}V_{a}}\bar{\Gamma}_{A}\Gamma_{A}\right)\nonumber\\
\\
\mathcal{L}_{W}&=&-\frac{\mu}{2}\int d\theta^{+}\sum_{A}\Gamma_{A}J^{A}+h.c.\\\
\mathcal{L}_{GS}&=&\frac{1}{4}\int d^{2}\theta^{+}\sum_{a,b,l}\hat{\mathcal{C}}^{[a}_{l}\mathcal{Q}^{b]l}V_{a}A_{b}-\frac{1}{4}\sum_{a,l}\hat{\CC}^{al} \int d\theta^{+}\Upsilon_{a} Y_{l}+h.c.\\\ \label{SFLagend}
\mathcal{L}_{FI}&=&-\frac{1}{8}\sum_{a}t^{a} \int d\theta^{+}\Upsilon_{a}+h.c.
\end{eqnarray}

Restricting to the case of $\hat{\mathcal{C}}=0$, i.e. the classically gauge invariant case, the component expansion in Wess-Zumino gauge can be organized as:

\begin{eqnarray}
\mathcal{L}&=&\mathcal{L}_{b}+\mathcal{L}_{f}+\mathcal{L}_{a}+\mathcal{L}_{i}\nonumber\\
\mathcal{L}_{b}&=&\sum_{a}\(\frac{1}{2e^{2}_{a}}F_{01a}^{2} - \frac{1}{2}\theta^{a} F_{01 a}\)+\sum_{I}D_{+}\bar{\phi}_{I}D_{-}\phi_{I}+\sum_{l}k^{2}_{l}D_{+}\bar{y}_{l}D_{-}y_{l}\\
\mathcal{L}_{f}&=&i\sum_{a}\frac{1}{2e^{2}_{a}}\bar{\lambda}_{-a}\partial_{+}\lambda_{-a}+i\sum_{I}\bar{\psi}_{+I}D_{-}\psi_{+I}+i\sum_{l}k^{2}_{l}\bar{\chi}_{+l}\partial_{-}\chi_{+l}+i\sum_{A}\bar{\gamma}_{-A}D_{+}\gamma_{-A}\\
&+&i\sum_{al}\frac{k^{2}_{l}\mathcal{Q}^{al}}{2}(\lambda_{-a}\chi_{+l}+\bar{\lambda}_{-a}\bar{\chi}_{+l})\nonumber\\
\mathcal{L}_{a}&=&\sum_{a}\frac{1}{2e^{2}_{a}}D^{2}_{a}+\sum_{a}\frac{D_{a}}{2}\left(\sum_{I}Q^{a}_{I}|\phi_{I}|^{2}+\sum_{l} \CQ^{al}k^{2}_{l}(y_{l}+\bar{y}_{l})- r_{a}\right)+\sum_{A}|G_{A}|^{2}+\frac{\mu}{\sqrt{2}}\sum_{A}(G_{A}J^{A}+h.c.)\nonumber\\
\\
\mathcal{L}_{i}&=&\frac{\mu}{\sqrt{2}}\sum_{A}\gamma_{-A}\left(\sum_{I}\partial_{I}J^{A}\psi_{+I}+\sum_{l}\partial_{l}J^{A}\chi_{+l}+h.c.\right)\
 -\sum_A\bar \gamma_{-A}  \(\sum_I \psi^I_+ \partial_I E_A  + \sum_{l}\chi_{+l}  \partial_{l} E_A + h.c\)\nonumber \\
&+&\frac{i}{2}\sum_{I,a}Q_{I}^{a}(\bar{\phi}_{I}\lambda_{-a}\psi_{+}+h.c.)-\sum_A | E_A|^2.  
\end{eqnarray}

The auxiliary Lagrangian can be put in a more standard form by solving for the auxiliary fields.
\begin{eqnarray}
\mathcal{L}_{a}&=&-\sum_{a}\frac{1}{2e^{2}}D^{2}_{a}-\sum_{A}|G_{A}|^{2} - \sum_A |E_A|^2\\
D_{a}&=&-\frac{e_{a}^{2}}{2}(\sum_{I}Q_{I}^{a}|\phi_{I}|^{2}+\sum_{l}\mathcal{Q}^{al}k^{2}_{l}(y_{l}+\bar{y}_{l})-r_{a})\\
G_{A}&=&-\frac{\mu}{\sqrt{2}}\bar{J}_{A}
\end{eqnarray}
Thus the bosonic potential becomes
\bea
U = \sum_{a}\frac{e^{2}_{a}}{8}\(\sum_{I}Q^{a}_{I}|\phi_{I}|^{2}+\sum_{a}\mathcal{Q}^{al}k^{2}_{l}(y_{l}+\bar{y}_{l}) - r_{a}\)^2 + \frac{\mu^{2}}{2}\sum_A |J_A|^2  + \sum_A |E_A|^2.
\eea


\section {OPEs}\label{opes}

We consider a complex scalar with action $\mathcal{L}_{b}=-c \, \partial_{\mu}\bar{\phi}\partial^{\mu}\phi=\frac{c}{2}(\partial_{+}\bar{\phi}\partial_{-}\phi+\partial_{-}\bar{\phi}\partial_{+}\phi)$, and a fermion with action, $\mathcal{L}_{f}=ic\, \bar{\psi}_{\pm}\partial_{\mp}\psi_{\pm}$. Also we consider the action of the form
\bea
S = \frac{1}{4\pi} \int d^2 x \;  \CL(x) 
\eea
Then the two point functions are:
\begin{eqnarray}
\langle\bar{\phi}(x)\phi(y)\rangle&=&\frac{1}{ c}\log(x-y)^{2}\\
\langle\bar{\psi}_{\pm}\psi_{\pm}\rangle&=&\frac{i}{c}\frac{1}{x^{\pm}-y^{\pm}}
\end{eqnarray}

\subsection{Operator Product Expansion with single anomalous $U(1)$  }

Let us list the currents and stress tensor that we want to calculate OPE's with

\bea
j_G^{+}  &=& - i \sum_I   Q_I   \phi_I \p_- \bar \phi_I  + \sum_A q_A \bar \gamma_{- A} \gamma_{- A } -  i k^2 \p_- \bar y  + i \f{\CN_G}{4\pi} \p_-y  - i \hat \CC\p_- y  \nonumber \\
j_L^+ &=& - i \sum_I \alpha^L_I \phi_I \p_- \bar \phi_I + \sum_A \beta^L_A \bar \gamma_{-A} \gamma_{-A} - i k^2 \kappa^L \p_-\bar y + i \f{\CN_L}{4\pi} \p_- y - i \kappa^L \hat \CC\p_- y  \nonumber \\
t^{++} &=&- \f 12 \(\sum_I \p_- \phi_I  \p_- \bar \phi_I + i \sum_A \bar \gamma_{-A} \p_- \gamma_{-A} + k^2 \p_- y \p_- \bar y + \f {i}{2e^2} \lambda \p_- \bar \lambda \)  \nonumber \\
&& + \f 14 \p_- \(\sum_I \alpha^R_I \phi_I \p_- \bar \phi_I + i \sum_A (\beta^R_A+1) \bar \gamma_{-A} \gamma_{-A} + k^2 \kappa^R \p_-\bar y-\f{\CN_R}{4\pi} \p_- y +\kappa^R \hat \CC \p_- y \) \nonumber \\
j_R^+ &=& -i \sum_I \alpha^R_I \phi_I \p_{-}\bar \phi_I +\sum_A \beta^R_A \bar \gamma_{-  A} \gamma_{- A}  -i  k^2 \kappa^R \p_{-}\bar y  + \frac{1}{2} \bar \lambda_{-} \lambda_{-}  - i \,\kappa^R \hat \CC\p_- y  \nonumber \\
j_R^- &=& i\sum_I \(\alpha_I^R \phi_I \p_{+} \bar \phi_I - i \(\alpha_I^R-1\)\bar \psi_{+ I } \psi_{+ I }\) + i\( k^2 \kappa^R \p_{+} y  + i k^2 \bar \chi_{+} \chi_{+}\)  - i  \, \kappa^R \hat \CC  \p_+ y_ . \
\eea

OPEs for the currents are: 

\bea
j_G^{+}(x) j_G^{+}(y) &\sim&  \frac{1}{(x^- -   y^-)^2} \( \sum_I  Q_I   Q_I  - \sum_A q_A  q_A - 2 \hat \CC \)   \nonumber \\
j_G^{+}(x) j_L^+(y) &\sim&   \frac{1}{(x^- -  y^-)^2} \(\sum_I  Q_I \alpha_I^L - \sum_A q_A   \beta_A^L  - 2 \hat \CC \kappa^L  \)  \nonumber\\
j_L^+(x) j_L^+(y) &\sim&  \frac{1}{(x^- -   y^-)^2} \(\sum_I \alpha_I^L  \alpha_I^L - \sum_A  \beta_A^L  \beta_A^LL - 2 \hat \CC  (\kappa^L )^2 \) \nonumber \\
j_R^+(x) j_R^+(y) &\sim&  \frac{1}{(x^- -   y^-)^2} \(\sum_I \alpha_I^R  \alpha_I^R - \sum_A  \beta_A^R  \beta_A^R - \underbrace{1}_{\lambda_- }- 2 \hat \CC  (\kappa^R )^2  \) \nonumber \\
j_R^-(x) j_R^-(y) &\sim&  \frac{1}{(x^+ -   y^+)^2} \(\sum_I \alpha_I^R  \alpha_I^R - \sum_I\(\alpha_I^R - 1\)^2- \underbrace{1}_{\chi_+}  \) \nonumber \\
j_G^{+}(x) t^{++} (y)&\sim&  \frac{i}{2(x^- - y^-)^3} \;  \(\sum_I Q_I  \alpha_I^R - \sum_I Q_I - \sum_A q_A \beta_A^R - 2  \hat \CC \kappa^R \) \nonumber \\
&&-\frac{1}{2} \frac{ j_G^{+}}{\(x^- - y^-\)^2}-\frac{1}{2} \frac{\p_- j_G^{+}}{\(x^- - y^-\)} \nonumber \\
j_L^+(x) t^{++}(y) &\sim&   \frac{i}{2(x^- - y^-)^3} \;  \(\sum_I \alpha_I^L \alpha_I^R - \sum_I \alpha^L_I - \sum_A \beta_A^L  \beta_A ^R   - 2 \hat \CC \kappa^L \kappa^R\)\nonumber \\
&&-\frac{1}{2}  \frac{j_L^{+}}{\(x^- - y^-\)^2} -\frac{1}{2}  \frac{\p_- j_L^{+}}{\(x^- - y^-\)} \nonumber \\
t^{++}(x) t^{++}(y) &\sim& \f{1}{8(x^- - y^-)^4}\;  \(\sum_I \(3 (\alpha^R_I -1)^2 -1 \)  + \sum_A \(1- 3 (\beta^{R }_A)^2\)  - 6 \hat \CC   (\kappa^R )^2  + \underbrace{2}_{y} -  \underbrace{2}_{\lambda_-} \) \nonumber \\
&&-\frac{1}{2}\(\frac{t^{++}}{\(x^- - y^-\)^2}+\frac{\p_- t^{++}}{\(x^- - y^- \) }\) . 
\eea




\subsection{Operator Product Expansion with multiple $U(1)$s  }
After rescaling $\U_a \rightarrow e_a^2 \U_a$ in the deep UV, the free operators we calculate OPEs are
\bea
j_G^{a +}  &=& - i \sum_I   Q^a_I   \phi_I \p_- \bar \phi_I  + \sum_A q^a_A \bar \gamma_{- A} \gamma_{- A } -  i\sum_l k_l^2 \CQ^a_l \p_- \bar y_l   - i \sum_l \CU^{al} \p_- y_l \nonumber \\
j_L^+ &=& - i \sum_I \alpha^L_I \phi_I \p_- \bar \phi_I + \sum_A \beta^L_A \bar \gamma_{-A} \gamma_{-A} - i \sum_l k_l^2 \kappa_l^L \p_-\bar y_l  - i \sum_l \CU^{L l } \p_- y_l  \nonumber\\
t^{++} &=&- \f 12 \(\sum_I \p_- \phi_I  \p_- \bar \phi_I + i \sum_A \bar \gamma_{-A} \p_- \gamma_{-A} + \sum_l k_l^2 \p_- y_l \p_- \bar y_l + \sum_a \f {i}{2} \lambda_a \p_- \bar \lambda_a \)   \nonumber \\
&& + \f 14 \p_- \(\sum_I \alpha^R_I \phi_I \p_- \bar \phi_I + i \sum_A (\beta^R_A+1) \bar \gamma_{-A} \gamma_{-A} + \sum_l k_l^2 \kappa_l^R \p_-\bar y_l +\CU^{R l} \p_- y_l  \) \nonumber \\
j_R^+ &=& -i \sum_I \alpha^R_I \phi_I \p_{-}\bar \phi_I +\sum_A \beta^R_A \bar \gamma_{-  A} \gamma_{- A}  -i  \sum_ l k_l^2 \kappa_l ^R \p_{-}\bar y_l  + \sum_a \frac{1}{2} \bar \lambda_{-a} \lambda_{-a}  -i \sum_l \CU^{R l }\p_- y_l  \nonumber \\
j_R^- &=& i\sum_I \(\alpha_I^R \phi_I \p_{+} \bar \phi_I - i \(\alpha_I^R-1\)\bar \psi_{+ I } \psi_{+ I }\) + i\sum_l\( k_l^2 \kappa_l^R \p_{+} y_l  + i k_l^2 \bar \chi_{l+} \chi_{l+}\)  -i \sum_l \CU^{R l }\p_+ y_l . \nonumber \\
\eea
Then the free OPEs of the currents are:
\bea
j_G^{a+}(x) j_G^{b+}(y) &\sim&  \frac{1}{(x^- -   y^-)^2} \( \sum_I  Q^a_I   Q^b_I  - \sum_A q^a_A  q^b_A - 2 \sum_l \CU_l^{(a}\CQ_l ^{b)}  \)   \nonumber \\
j_G^{a+}(x) j_L^+(y) &\sim&   \frac{1}{(x^- -  y^-)^2} \(\sum_I  Q^a_I \alpha_I^L - \sum_A q^a_A   \beta_A^L - \sum_l \(\CU^{al} \kappa^L_l + \CU^{L l } \CQ^a_l \) \)  \nonumber\\
j_L^+(x) j_L^+(y) &\sim&  \frac{1}{(x^- -   y^-)^2} \(\sum_I \alpha_I^L  \alpha_I^L - \sum_A  \beta_A^L  \beta_A^L - 2\sum_l  \CU^{L l } \kappa^L_l  \) \nonumber \\
j_R^+(x) j_R^+(y) &\sim&  \frac{1}{(x^- -   y^-)^2} \(\sum_I \alpha_I^R  \alpha_I^R - \sum_A  \beta_A^R  \beta_A^R - \sum_a 1 - 2\sum_l  \CU^{R l } \kappa^R_l  \) \nonumber \\
j_R^-(x) j_R^-(y) &\sim&  \frac{1}{(x^+ -   y^+)^2} \(\sum_I \alpha_I^R  \alpha_I^R- \sum_I \(\alpha_I^R - 1\)^2-\sum_l 1  \) \nonumber \\
j_G^{a+}(x) t^{++} (y)&\sim&  \frac{i}{2(x^- - y^-)^3} \;  \(\sum_I Q^a_I  \alpha_I^R - \sum_I Q^a_I - \sum_A q^a_A \beta_A^R - \sum_l\(  \CU^{al} \kappa^R_l + \CU^{R l} \CQ^a_l \)\) \nonumber \\
&&-\frac{1}{2} \frac{ j_G^{a+}}{\(x^- - y^-\)^2}-\frac{1}{2} \frac{\p_- j_G^{a+}}{\(x^- - y^-\)} \nonumber \\
j_L^+(x) t^{++}(y) &\sim&   \frac{i}{2(x^- - y^-)^3} \;  \(\sum_I \alpha_I^L \alpha_I^R - \sum_I \alpha^L_I - \sum_A \beta_A^L  \beta_A ^R   - \sum_l\(  \CU^{L l} \kappa^R_l + \CU^{R l} \kappa^L_l \) \)\nonumber \\
&&-\frac{1}{2}  \frac{j_L^{+}}{\(x^- - y^-\)^2} -\frac{1}{2}  \frac{\p_- j_L^{+}}{\(x^- - y^-\)} \nonumber 
\eea
\bea
t^{++}(x) t^{++}(y) &\sim& \f{1}{8(x^- - y^-)^4}\;  \(\sum_I \(3 (\alpha^R_I -1)^2 -1 \)  + \sum_A \(1- 3 (\beta^{R }_A)^2\) - 6 \sum_l \CU^{Rl} \kappa^R_l  + \sum_l 2 - \sum_a 2 \) \nonumber \\
&&-\frac{1}{2}\(\frac{t^{++}}{\(x^- - y^-\)^2}+\frac{\p_- t^{++}}{\(x^- - y^- \) }\) . 
\eea

\section{Quantum Chirality}\label{GI Q Chirality}
As mentioned above, classically chiral operators can be anomalous. In this section we check that chirality holds within correlation functions. If we only focus on supersymmetric vacua then:
\bea
\<\[ {\DB}_+, J(x)\] \CO (y)\> = - \<J(x)\,  \[ {\DB}_+,  \CO (y)\]_\pm \>.
\eea
In order to verify chirality we want an operator $\CO(y)$, where the action of $\DBp$ is known, following \cite{AdamsLapan:2009}, \cite{Silverstein:1994},  we choose $\CO(y) = \p_- \bar \U(y)$.  

Using the equations of motion $\DBp\partial_{-}\bar{\Upsilon} =  2 i e^2 J_G^+ (y)$.  A necessary condition for quantum chirality is:
\bea
\<\DBp J(x) \;  \p_- \bar \U(y) \> &\approx& -  \<J(x) \;  \DBp \p_- \bar \U(y) |_{EOMs}\> = -  2i e^2 \<J(x) J_G^+(y)  \>=0.
\eea

Thus quantum chirality requires the vanishing of the leading singular part of the $J \, J_G^+$ OPE. Let us first investigate $T^{++}$. The lowest component of $T^{++}$ and $J_{G}^{+}$ give:

\bea
j_G^+(x) \,  t^{++} (y)&\sim&   \f{i}{2(x^- - y^-)^3}  \;  \(\sum_I Q_I \alpha_I^R - \sum_I Q_I - \sum_A q_A \beta_A^R \)-\frac{1}{2} \frac{ j_G^{+}}{\(x^- - y^-\)^2}-\frac{1}{2} \frac{\p_- j_G^{+}}{\(x^- - y^-\)}.  \nonumber \\
\eea
Therefore requiring $ \DBp T^{++} =0$, relies on having:
\bea
 \sum_I Q_I \alpha_I^R - \sum_I Q_I - \sum_A q_A \beta_A^R  = 0.
\eea 
This condition is equivalent to a particular R-symmetry being non-anomalous.  Thus quantum chirality of the left moving stress tensor requires a non-anomalous R-symmetry.

Next let's consider $J_L^+$, which has an OPE with the gauge current:
\bea
j_G^+(x) \,   j_L^+(y) &\sim&  \frac{1}{(x^- -  y^-)^2} \(\sum_I Q _I  \alpha_I^L - \sum_A q_A \beta_A^L \).  \nonumber
\eea
Thus non-chirality of $U(1)_{L}$ boils down to the vanishing of an L-anomaly:
\bea
\DBp J_L^+ \propto  \(\sum_I Q _I  \alpha_I^L - \sum_A q_A \beta_A^L \) \U.
\eea

For the gauge current  $J_G^+$ to be a quantum chiral operator, the same analysis gives:
\bea
\DBp J_G^+ \propto \(\sum_I Q_I  Q_I -  \sum_A q_A q_A\) \U.
\eea

In short, the quantum chirality of a current requires the various anomaly coefficients vanish. The existence of a non-anomalous gauge current, L current and R current is necessary for quantum chirality of $J_L^+$ and $T^{++}$, and tells us that the lowest components of these operators are elements of cohomology, which become the $U(1)_L$ current and left moving stress tensor for the IR CFT. 

\bibliographystyle{JHEP}
\bibliography{masterbib}

\end{document}